\documentclass[acmsmall,screen,nonacm]{acmart}

\setcopyright{none}

\renewcommand\footnotetextcopyrightpermission[1]{}

\AtBeginDocument{%
}

\usepackage{capt-of}
\usepackage{amsmath}
\usepackage{mathrsfs}
\usepackage{amsthm}

\usepackage{graphicx}
\usepackage{wrapfig}
\usepackage{booktabs}
\usepackage{colortbl}
\usepackage{tabularx}
\usepackage{array}
\usepackage{float}
\usepackage{subfig}
\usepackage{listings}
\usepackage{upquote}
\usepackage[most]{tcolorbox}

\usepackage{algorithm}
\usepackage{algorithmic}

\usepackage{tikz}
\usepackage{fontawesome5}
\usepackage{pifont}

\usepackage{xcolor}
\usepackage{microtype}
\usepackage{underscore}
\usepackage{needspace}
\usepackage{enumitem}
\usepackage[disable]{todonotes}
\usepackage{xspace}

\usetikzlibrary{positioning}
\usetikzlibrary{tikzmark}

\makeatletter
\@ifundefined{definition}{%
  \theoremstyle{definition}
  \newtheorem{definition}{Definition}
}{}
\@ifundefined{example}{%
  \theoremstyle{definition}
  
}{}
\makeatother

\definecolor{cmt}{gray}{0.42}
\definecolor{dred}{RGB}{170,35,35}
\definecolor{dgreen}{RGB}{30,120,50}

\definecolor{diffaddbg}{RGB}{230,255,236}
\definecolor{diffdelbg}{RGB}{255,235,233}
\definecolor{diffemptybg}{RGB}{235,237,240}

\definecolor{codebg}{RGB}{248,248,248}
\definecolor{framegray}{RGB}{100,100,100}
\definecolor{numgray}{RGB}{120,120,120}

\definecolor{specgray}{HTML}{F4F5F7}
\definecolor{specline}{HTML}{D9DCE1}
\definecolor{nextgreen}{HTML}{28765F}
\definecolor{nextfill}{HTML}{EDF7F3}
\definecolor{lastred}{HTML}{A64B4B}
\definecolor{lastfill}{HTML}{FFF1F0}

\lstdefinestyle{reaperbug}{
  language=SQL,
  basicstyle=\ttfamily\scriptsize,
  keywordstyle=\bfseries\color{black},
  identifierstyle=\color{black},
  commentstyle=\color{cmt},
  stringstyle=\color{black},
  numbers=left,
  numberstyle=\normalfont\tiny\color{numgray},
  stepnumber=1,
  numbersep=7pt,
  xleftmargin=15pt,
  frame=none,
  columns=fullflexible,
  keepspaces=true,
  breaklines=true,
  breakatwhitespace=false,
  showstringspaces=false,
  upquote=true,
  captionpos=t,
  aboveskip=4pt,
  belowskip=4pt,
  escapeinside={(*@}{@*)}
}

\lstdefinestyle{case}{
  basicstyle=\ttfamily\scriptsize\color{black},
  identifierstyle=\color{black},
  keywordstyle=\bfseries\color{black},
  commentstyle=\color{cmt},
  stringstyle=\color{black},
  numbers=left,
  numberstyle=\normalfont\tiny\color{numgray},
  stepnumber=1,
  numbersep=7pt,
  xleftmargin=15pt,
  frame=none,
  columns=fullflexible,
  keepspaces=true,
  showstringspaces=false,
  breaklines=true,
  breakatwhitespace=false,
  upquote=true,
  language={},
  captionpos=t,
  aboveskip=4pt,
  belowskip=4pt,
  escapeinside={(*@}{@*)}
}

\lstdefinestyle{sqlcase}{
  language=SQL,
  basicstyle=\ttfamily\footnotesize,
  keywordstyle=\color{blue!65!black},
  commentstyle=\color{gray!75!black}\itshape,
  stringstyle=\color{green!40!black},
  columns=fullflexible,
  keepspaces=true,
  showstringspaces=false,
  breaklines=true,
  breakatwhitespace=false,
  frame=single,
  rulecolor=\color{black!18},
  framesep=3pt,
  xleftmargin=2pt,
  xrightmargin=2pt,
  aboveskip=4pt,
  belowskip=4pt,
  numbers=none
}

\begin{document}

\title{Fix One, Miss Another: Leveraging Historical Bug Fixes to Uncover Residual Bugs in Database Systems}

% -----------------------------------------------------------------------------
% -----------------------------------------------------------------------------

\author{Yankai Rong}
\email{yankairong@ruc.edu.cn}
\affiliation{%
  \institution{Renmin University of China}
  \city{Beijing}
  \country{China}
}

\author{Zhaoyang Zhang}
\email{zhaoyangzhang@ruc.edu.cn}
\affiliation{%
  \institution{Renmin University of China}
  \city{Beijing}
  \country{China}
}

\author{Shuang Liu}
\email{Shuang.Liu@ruc.edu.cn}
\affiliation{%
  \institution{Renmin University of China}
  \city{Beijing}
  \country{China}
}

\author{Wei Lu}
\email{lu-wei@ruc.edu.cn}
\affiliation{%
  \institution{Renmin University of China}
  \city{Beijing}
  \country{China}
}

\author{Xiaoyong Du}
\email{duyong@ruc.edu.cn}
\affiliation{%
  \institution{Renmin University of China}
  \city{Beijing}
  \country{China}
}

% -----------------------------------------------------------------------------
% Abstract
% -----------------------------------------------------------------------------

\begin{abstract}

Database management systems (DBMSs) implement SQL constructs through grammar-driven dispatch handlers. Although handlers for syntactic siblings implement different SQL constructs, they often share a common implementation pattern, including argument extraction, type checking, and permission validation. Consequently, a bug fix that hardens the handler exercised by a historical bug may leave the corresponding logic in sibling handlers vulnerable to the same semantic flaw. We refer to such defects as sibling bugs. Despite the structural similarity among sibling handlers, existing bug-finding techniques rarely exploit historical fixes to systematically uncover these related defects.

We present \textsc{Reaper}, a novel approach for detecting sibling bugs guided by historical bug fixes. \textsc{Reaper} first analyzes a bug fix to infer a trigger profile, which characterizes the conditions under which the defect is exposed, and a validation profile, which captures the semantic evidence needed to distinguish buggy behavior from correct behavior. It then projects the trigger profile onto syntactic siblings through grammar-guided test mutation, generating tests that exercise corresponding semantic paths in sibling handlers. Finally, \textsc{Reaper} validates suspicious behaviors using the inferred validation profile through a staged validation process, reducing false positives caused by complex DBMS behavior.

We evaluate \textsc{Reaper} on four widely used DBMSs—MariaDB, MySQL, OceanBase, and CockroachDB. \textsc{Reaper} uncovers 61 previously unknown bugs, of which 57 have been confirmed by developers and 13 have been fixed at the time of writing. These results demonstrate that historical bug fixes contain actionable semantic knowledge that can be systematically transferred across syntactic siblings to uncover otherwise overlooked defects.

\end{abstract}

% -----------------------------------------------------------------------------
% ACM Computing Classification System
% -----------------------------------------------------------------------------

%   <concept_id>10011007.10011006.10011008</concept_id>
%   <concept_desc>Software and its engineering~Software testing and debugging</concept_desc>
%   <concept_id>10002951.10002952.10003197</concept_id>
%   <concept_desc>Information systems~Database management system engines</concept_desc>

% \ccsdesc[500]{%
%   Software and its engineering~Software testing and debugging%
% }

% \ccsdesc[300]{%
%   Information systems~Database management system engines%
% }

% \keywords{%
%   DBMS,
%   database testing,
%   test-case generation
% }

\maketitle

% -----------------------------------------------------------------------------
% Main paper
% -----------------------------------------------------------------------------

\section{Introduction}
\label{sec:introduction}

\begin{figure}[t]
  \centering
  \vspace{-5mm}
  \includegraphics[width=\linewidth]{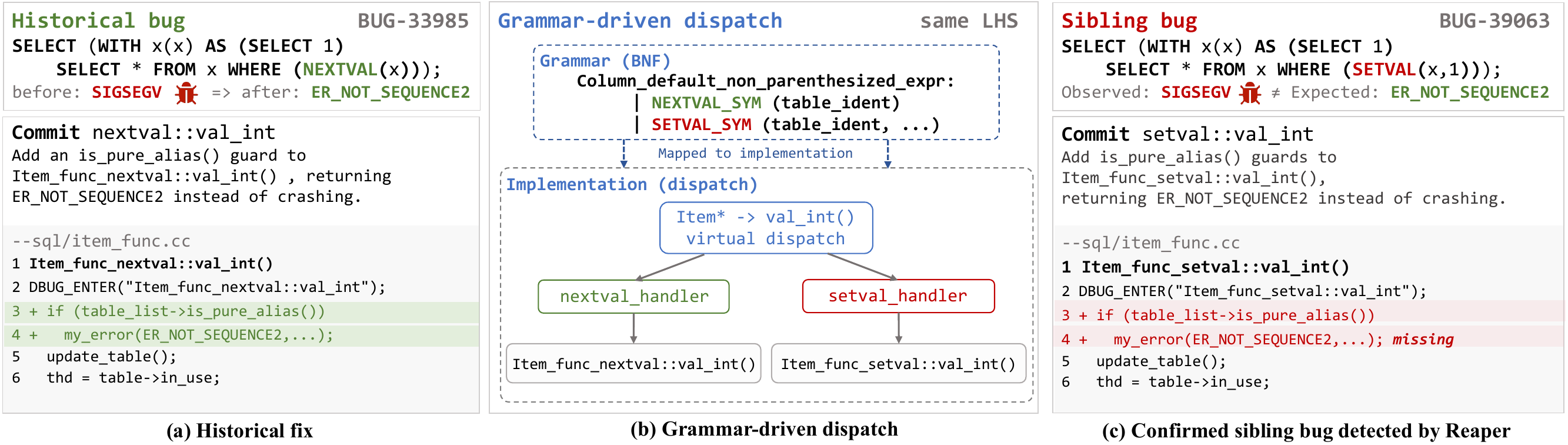}
  \vspace{-6mm}
  \caption{Reaper Reveals an Unpropagated Fix in MariaDB.}
  \vspace{-5mm}
  \label{fig:motivating-examples}
\end{figure}

The Structured Query Language (SQL) is a declarative language whose behavior is
specified in the ISO/IEC 9075 standard~\cite{sql-standard}. Each syntactic
element/construct of the standard is defined through a format expressed as a BNF
production, accompanied by the Syntax Rules, Access Rules, and General Rules
that govern its processing. 
Modern DBMSs~\cite{mysql_doc,mariadb_doc,oceanbase_doc,cockroachdb_doc} implement these specifications through
a grammar-driven architecture: a recognized SQL construct is routed by a
dispatcher to a dedicated handler responsible for its execution. This recurring
architectural pattern is what we call the handler dispatch pattern~\cite{sqlbull,duckdb,sqlancerpp,tlp,norec,pqs}. 
% A dispatcher selects one handler from a set of sibling handlers based on a key, such as a SQL
% token type or a parse-tree node tag. Because the sibling handlers implement
% syntactically analogous constructs, their implementations share a substantial
% common subsequence---argument extraction, type checking, and permission
% validation---that is logically orthogonal to each handler's own distinctive
% logic~\cite{sqlbull,duckdb,sqlancerpp,tlp,norec,pqs}.

The handler dispatch structure is directly mirrored by a grammar's production
rule: the left-hand-side non-terminal of a production rule corresponds to the dispatcher, and its right-hand-side alternatives correspond to the construct-specific handlers. Figure~\ref{fig:motivating-examples}(b) illustrates
this correspondence through the MariaDB sequence functions \texttt{NEXTVAL} and
\texttt{SETVAL}. These two constructs originate from the same left-hand-side non-terminal 
in the BNF grammar and share similar semantics---both resolve a sequence object
and then apply a form-specific effect. In the implementation, the virtual
function \texttt{Item::val\_int()} serves as the dispatcher; the functions
\texttt{Item\_func\_nextval::val\_int()} and
\texttt{Item\_func\_setval::val\_int()} are the two sibling
handlers responsible for \texttt{NEXTVAL} and \texttt{SETVAL}, respectively.

Because sibling handlers share substantial common logic, they should evolve
together: a bug uncovered in one handler is a strong signal that an analogous
bug may exist in its siblings. In practice, however, developers fix a system
based on a reported bug, which typically pinpoints only one handler. A fix
applied to one handler therefore often leaves its siblings exposed to the same
bug---a phenomenon we term a \textit{sibling bug}.

Figure~\ref{fig:motivating-examples}(c) illustrates a sibling bug we identified
in MariaDB, with respect to a historical bug shown in 
Figure~\ref{fig:motivating-examples}(a). The constructs \texttt{NEXTVAL} and
\texttt{SETVAL} originate from the same BNF production and both require their
argument to resolve to a valid sequence object. The original bug, BUG~\#33985, was
fixed by adding an \texttt{is\_pure\_alias()} guard only to the \texttt{NEXTVAL}
handler; the accompanying regression test exercised only this construct. The
sibling handler \texttt{SETVAL} was therefore left unfixed despite sharing an
analogous implementation. Replacing \texttt{NEXTVAL} in the original
bug triggering query with \texttt{SETVAL} therefore exercised analogous processing logic in the sibling handler and still triggered a server crash. We reported this sibling bug as BUG~\#39063,
whose follow-up fix propagated the same guard to the \texttt{SETVAL} handler.

However, systematically detecting sibling bugs remains non-trivial. Existing
DBMS fuzzers explore syntactic spaces or execution
paths~\cite{liang2025parser,squirrel,dynsql,polyglot,qpg,sqlaser,sqlancerpp,DBcover},
but cannot exercise sibling handlers that may remain inconsistent
after repairing the reported bugs. Bridging this gap requires overcoming three
challenges:

\textbf{First, precisely recovering the bug-triggering and bug-validation conditions.}
Historical fix records often contain noisy and lengthy context, making it difficult to recover the bug-triggering conditions and the semantic evidence used to distinguish buggy behavior from correct behavior. The challenge is thus to distill a historical fix into a concise and actionable profile that captures both \emph{when} the bug can be triggered and \emph{how} its manifestation can be validated.
% especially the bug-triggering construct, and the corresponding bug-validation profiles needed for subsequent sibling-bug detection.

\textbf{Second, systematically exercising sibling handlers while preserving bug-triggering conditions.}  
% Blindly mutating the original query does not systematically reach
% sibling handlers associated with another alternative. Moreover, mutating the original query to exercise a sibling alternative requires adapting the query to the corresponding construct while preserving the bug-triggering conditions of the historical bug-triggering query. 
Blindly mutating a historical bug-triggering query does not systematically reach handlers implementing syntactic siblings of the original construct. More importantly, simply replacing the original construct with a sibling alternative may invalidate the conditions that expose the historical defect. Effective sibling testing therefore requires simultaneously satisfying two constraints: the generated query must be routed to the intended sibling handler, while preserving the relevant semantic conditions under which the historical bug is triggered. The challenge is to transform a historical bug-triggering query across syntactic siblings without losing its defect-triggering characteristics.

\textbf{Third, accurately validating sibling bugs without ground truth.}
% A sibling query is generally not semantically equivalent to the original one
% and lacks a known expected result. Existing DBMS testing oracles relying on
% equivalence-based metamorphic relations~\cite{tlp,norec,dqe} or approximation
% relations~\cite{ValScope} cannot be directly applied to validating sibling bugs.
Unlike conventional regression testing, a sibling query generally has no known expected result. It is also not necessarily semantically equivalent to the historical bug-triggering query, making direct result comparison inappropriate. Existing DBMS testing oracles based on equivalence-oriented metamorphic relations~\cite{tlp,norec,dqe} or approximation relations~\cite{ValScope} therefore cannot be directly applied to determine whether a suspicious sibling execution exhibits the same semantic defect. The challenge is to derive validation evidence from the historical fix and determine whether the sibling execution violates the semantic property associated with the original bug, without a predefined ground-truth result.

To address these challenges, we present \textsc{Reaper}, a DBMS fuzzing framework that turns historical bug fixes into reusable probes for detecting sibling bugs. \textsc{Reaper} follows a \emph{profile--generate--validate} pipeline consisting of three phases.
\textbf{Bug Profile Construction} addresses the challenge of recovering actionable triggering and validation conditions from noisy fixes. We design a multi-agent system consisting of a \emph{Query Analyzer}, a \emph{Patch Analyzer} and an \emph{Orchestrator}. The \emph{Query Analyzer} identifies the bug-triggering construct and historical failure pattern, while the \emph{Patch Analyzer} infers the correctness requirement from the fix. The \emph{Orchestrator} reconciles these complementary findings into a \emph{Bug Profile} containing trigger conditions and validation evidence. 
\textbf{Sibling Test Generation} addresses the challenge of reaching sibling handlers while preserving bug-triggering conditions. \textsc{Reaper} identifies sibling constructs from the SQL grammar and transforms historical bug-triggering queries to exercise those sibling handlers. A \emph{Concretizer} then satisfies sibling-specific semantic constraints while preserving the relevant trigger conditions in the \emph{Bug Profile}. 
\textbf{Sibling Bug Validation} addresses the challenge of validating sibling bugs without ground truth. \textsc{Reaper} validates sibling-bug candidates through a three-stage
checking process: failure-pattern matching, criterion applicability
checking, and fix-derived oracle checking. These stages reuse semantic evidence from the historical fix to determine whether sibling executions violate the corresponding correctness requirement. 
We evaluated \textsc{Reaper} on four widely used production DBMSs—MariaDB, MySQL, OceanBase, and CockroachDB. It uncovered 61 previously unknown bugs, including 50 logic bugs and 11 crash/fatal-error bugs; 57 have been confirmed and 13 have been fixed by upstream developers.

In summary, this paper makes the following contributions:

\begin{itemize}

    \item \textbf{Insight.}
    We identify \emph{sibling bugs}, a class of residual DBMS bugs that arise
    when a localized fix corrects one implementation while a related
    \emph{sibling handler} remains vulnerable to the same semantic flaw. We further observe 
    that historical bug fixes encode actionable knowledge about both bug-triggering conditions and correctness requirements, which can serve as semantic anchors for systematically examining sibling implementations.

    \item \textbf{Methodology.}
    We design and implement \textsc{Reaper}, a fix-guided DBMS fuzzing framework following a profile--generate--validate pipeline.
    \textsc{Reaper} constructs a Bug Profile from historical bug fix reports, identifies sibling constructs through a grammar relation graph, and generates sibling tests while preserving the historical triggering conditions. It then validates suspicious executions using fix-derived validation evidence, without requiring ground-truth results. 
%     \textsc{Reaper} uses collaborating LLM agents to construct a Bug Profile
% from historical bug fixes. It then employs a grammar relation graph to identifies sibling constructs, and generates sibling tests while preserving the historical triggering conditions. For validation, it uses the Validation Profile to match the historical failure pattern, check fix-derived criterion
%     applicability, and detect oracle violations without requiring ground-truth results.
    \item \textbf{Real-world impact.}
    We evaluate \textsc{Reaper} on MySQL, MariaDB, OceanBase, and CockroachDB. It uncovers $61$ previously unknown bugs, including $50$ logic bugs and $11$ crash or fatal-error bugs. Among them, $57$ have been confirmed and $13$ have been fixed by upstream developers. These results demonstrate that incomplete repair propagation remains a practical maintenance challenge in real-world DBMSs and that \textsc{Reaper} provides a way to systematically identify such residual bugs.

\end{itemize}

\section{Preliminaries}
\label{sec:backend}

\subsection{Grammar-Driven SQL Processing in DBMSs}
Modern DBMSs typically process an SQL query through several stages,
including parsing, semantic analysis, optimization, and execution.
After receiving a query, the DBMS first parses the input according to
its SQL grammar to determine its syntactic structure and creates an
internal representation for each recognized SQL construct. 
% The representation is then processed in subsequent stages, where the DBMS
% resolves database objects, checks semantic constraints, performs query
% rewriting and optimization, and eventually executes the resulting plan.
Throughout this process, different SQL constructs may be routed to
different processing paths according to handler-selection information
encoded in their internal representations, e.g., construct types,
expression classes, or function/operator identifiers.

SQL grammars provide a structural bridge between SQL constructs and
these processing paths. ISO/IEC~9075 specifies SQL syntactic elements
using BNF-style productions, while practical DBMSs maintain dialect
grammars to describe their supported SQL constructs~\cite{sql-standard}.
A production rule associates a left-hand-side nonterminal with one or
more right-hand-side alternatives, each defining a distinct syntactic
form of that nonterminal. As shown in Figure~\ref{fig:motivating-examples}(b), in MariaDB, the grammar alternative for
the SQL construct \texttt{NEXTVAL} is recognized during parsing and mapped to an
\texttt{Item\_func\_nextval} object in the internal representation. During
evaluation, the object is dispatched through the \texttt{Item::val\_int()}
interface to its handler \texttt{Item\_func\_nextval::val\_int()}, which implements the
sequence operation. This example illustrates the correspondence between grammar alternatives
and construct-specific processing paths, where sibling alternatives may lead
to distinct handlers implementing related semantics. 

\subsection{Syntactic Siblings and Handler Dispatch}

\begin{definition}[Syntactic Sibling]
\label{def:syntactic-sibling} 
We refer to SQL constructs derived from different alternatives of the same left-hand-side nonterminal as \emph{syntactic siblings}.
\end{definition}
% At the grammar level, a left-hand-side nonterminal may admit multiple alternative derivations. 
% We refer to SQL constructs derived from different alternatives of the same left-hand-side nonterminal as \emph{syntactic siblings}
Although syntactic siblings denote different SQL constructs, their shared grammatical role often leads them to enter the same processing framework before diverging into construct-specific behavior. Figure~\ref{fig:motivating-examples}(b) illustrates \texttt{NEXTVAL}
and \texttt{SETVAL} as syntactic siblings in MariaDB. Their grammar
alternatives, \texttt{NEXTVAL\_SYM '(' table\_ident ')'} and
\texttt{SETVAL\_SYM '(' table\_ident ',' ... ')'}, share the
same left-hand-side nonterminal. Here, \texttt{NEXTVAL\_SYM} and \texttt{SETVAL\_SYM} are grammar
tokens representing the SQL keywords \texttt{NEXTVAL} and
\texttt{SETVAL}, respectively.
At the implementation level, this grammar structure is reflected by the \emph{handler dispatch pattern}. A \emph{handler} is the implementation selected by a dispatch
decision to process a particular SQL construct. It may be realized
as a dedicated function or method, or as a distinct branch within
a shared implementation.
% We refer to handlers implementing syntactic siblings as \emph{sibling handlers}. 
\begin{definition}[Sibling Handler]
\label{def:sibling-handler} 
\emph{Sibling handlers} are handlers implementing syntactic siblings.
\end{definition}
As shown in Figure~\ref{fig:motivating-examples}(b), \texttt{NEXTVAL} and \texttt{SETVAL} are represented by \texttt{Item\_func\_nextval} and \texttt{Item\_func\_setval} within the same expression-processing framework. They share common preparation and evaluation interfaces. During evaluation, calls through the shared virtual interface \texttt{Item::val\_int()} are dispatched to \texttt{Item\_func\_nextval::val\_int()} or \texttt{Item\_func\_setval::\allowbreak val\_int()}, selecting the handler of the corresponding construct. 
Sibling handlers may further reuse preparation logic and auxiliary routines for object resolution, argument processing, and validation, while retaining separate execution paths for construct-specific semantics.

\subsection{Sibling Bugs}
\label{sec:prelim-sibling-bugs}

In a typical DBMS development workflow, a reported bug is reproduced by a failing query and localized to an implicated processing path. Developers patch that path and add a regression test to check that the historical failure does not recur. The patch therefore determines which implementation path is updated, while the regression test determines which triggering instance is exercised after the fix. Because both are typically centered on the reported construct, neither necessarily covers the construct-specific paths associated with its syntactic siblings.

This distinction matters when sibling handlers follow different execution paths but are subject to the same correctness requirement. A repair confined to the reported path may correct the historical violation without updating the corresponding logic in a sibling path. The original regression test can therefore pass while another grammar alternative still reaches an implementation that violates the same requirement. In the motivating example in Figure~\ref{fig:motivating-examples},
the repaired \texttt{NEXTVAL} handler rejects an invalid sequence
reference, whereas the sibling \texttt{SETVAL} handler still crashes
on the same reference. Both handlers should reject such references,
but the historical fix corrects only the \texttt{NEXTVAL} path. We formally define such residual bugs as follows:

\begin{definition}[Sibling Bug]
\label{def:sibling-bug} 
Two bugs are \emph{sibling bugs} if they occur in sibling handlers
and share the same bug-triggering conditions.
% A \emph{sibling bug} is a residual bug that violates a common correctness requirement in one sibling handler after a historical fix has corrected a violation of that requirement in another.
\end{definition}

Figure~\ref{fig:motivating-examples} illustrates how a localized fix
leaves a sibling bug unresolved in MariaDB. The historical query calls
\texttt{NEXTVAL(x)}, where \texttt{x} does not refer to a sequence
object. Instead of rejecting this invalid reference, the handler
attempts to access it as a sequence object and crashes. The historical
fix adds a check to reject the reference before sequence-object access,
and the accompanying regression test verifies the repaired
\texttt{NEXTVAL} path. However, the sibling \texttt{SETVAL} handler lacks the corresponding
check. Replacing \texttt{NEXTVAL(x)} with \texttt{SETVAL(x,1)} retains
the invalid reference while directing execution to this sibling handler.
Although both constructs require a valid sequence reference, the
\texttt{SETVAL} path still attempts sequence-object access and crashes,
as shown in Figure~\ref{fig:motivating-examples}(c).

\section{Approach}
\label{3-method}

\begin{figure}[t]
  \centering
  \includegraphics[width=\linewidth]{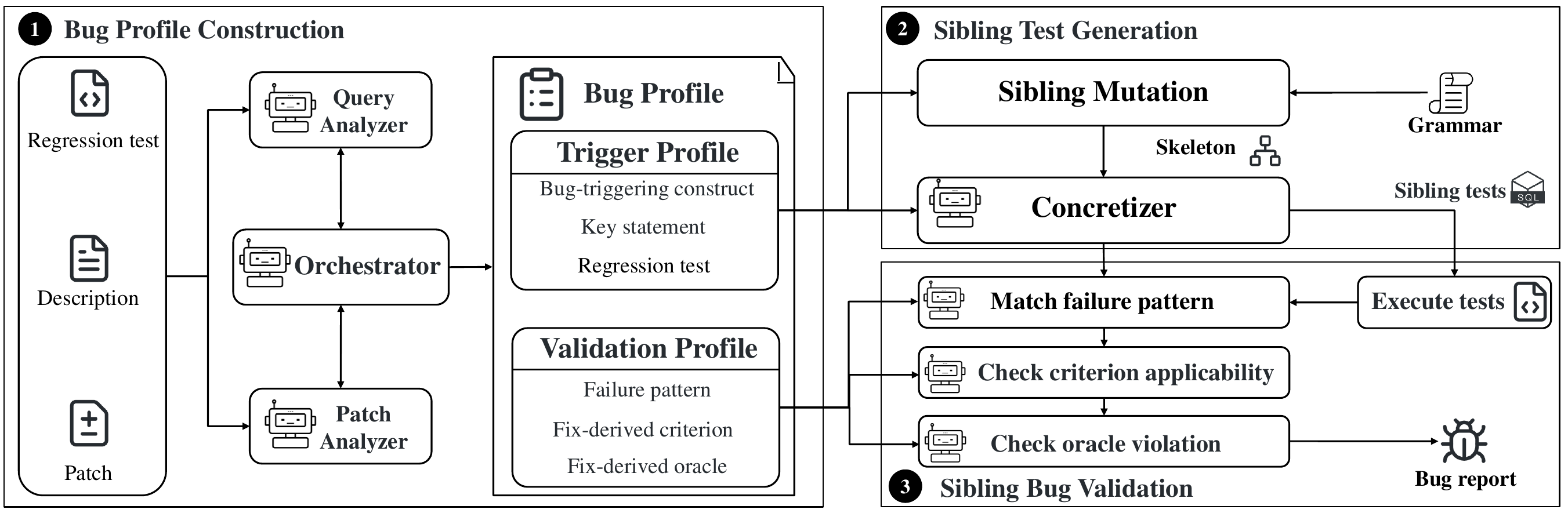}
  \caption{\textbf{Overview of \textsc{Reaper}.}}
  \label{fig:overview}
  \vspace{-0.8em}
\end{figure}

Figure~\ref{fig:overview} provides an overview of \textsc{Reaper}, which
consists of three components---Bug Profile Construction, Sibling Test
Generation, and Sibling Bug Validation. Given a historical fix,
\textsc{Reaper} generates and executes sibling tests to uncover sibling
bugs. During Bug Profile Construction, three collaborating LLM agents
analyze the regression test, bug description, and patch, integrating
query-side and repair-side evidence into a \emph{Bug Profile} comprising
a \emph{trigger profile} and a \emph{validation profile}
(\S\ref{sec:bug-profiling}). Then, the trigger profile guides Sibling Test Generation to produce
sibling test cases that preserve the bug-triggering conditions
(\S\ref{sec:mutation}). Sibling Bug Validation then applies the validation
profile in three stages. A sibling bug is reported only when the execution
matches the failure pattern, the fix-derived criterion applies to the
selected sibling, and the corresponding oracle, adapted to the sibling
test, is violated (\S\ref{sec:oracle}).

% Figure~\ref{fig:overview} illustrates the workflow of
% \textsc{Reaper}. Given a historical fix, \textsc{Reaper} follows a
% profile--project--validate pipeline that constructs and evaluates
% queries targeting sibling handlers.
% In Phase~\textbf{1} (\textit{Bug Profiling}), it recovers the
% information needed to reproduce the bugive behavior on sibling
% handlers and to recognize the same failure elsewhere, recording both as
% an \textit{Audit Blueprint} that guides the later phases;
% in Phase~\textbf{2} (\textit{Grammar-Guided Projection}), it
% identifies syntactic siblings based on the query that triggers the
% historical bug and projects the bug-trigger information onto the
% corresponding sibling handlers, generating queries that preserve the
% historical triggering conditions while adapting the surface syntax
% required to reach each sibling handler;
% in Phase~\textbf{3} (\textit{Sibling Bug Validation}), it executes the
% projected queries on the target DBMS and checks whether they reproduce
% the same failure as the historical fix, where a match indicates a
% residual sibling bug.
% In the following subsections, we present each phase in detail,
% including bug profiling (\S\ref{sec:bug-profiling}), grammar-guided
% projection (\S\ref{sec:mutation}), and sibling bug validation
% (\S\ref{sec:oracle}).
\subsection{Bug Profile Construction}
\label{sec:bug-profiling}

\begin{figure}[t]
  \centering
  \includegraphics[
    width=\linewidth,
  ]{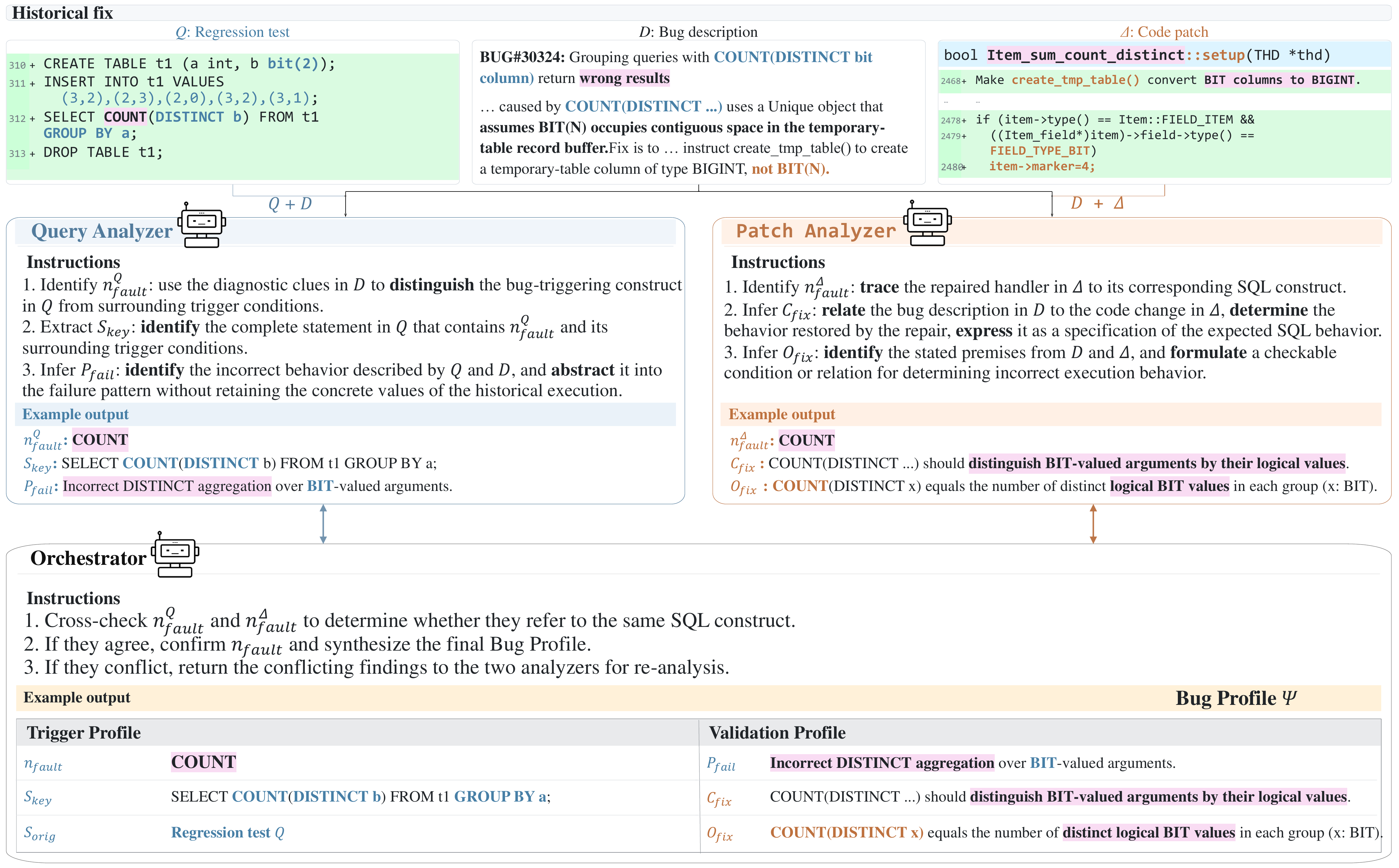}
  \caption{Bug Profile Construction.}
  \label{fig:bug-profiling}
  \vspace{-4mm}
\end{figure}

Turning fixes into probes requires lifting \textbf{path-local repair evidence}
into \textbf{sibling-oriented bug knowledge}. Our key insight is that
historical fixes encode bug knowledge in a path-specific form while sibling
testing requires transforming this local evidence into reusable
sibling-oriented knowledge for Sibling Test Generation and Sibling Bug
Validation. To support this transition, \textsc{Reaper} organizes the historical evidence
into a structured \emph{Bug Profile} $\Psi$. Before describing how $\Psi$ is
constructed, we first define what it contains and why each part is needed for
sibling testing.

\begin{definition}[Bug Profile]
\label{def:bug-profike} 
We define the \emph{Bug Profile} as
$\Psi=\langle T,V\rangle$, where $T$ is the \emph{trigger profile} and
$V$ is the \emph{validation profile}. The trigger profile is defined as
$T=\langle n_{\mathit{fault}},S_{\mathit{key}},S_{\mathit{orig}}\rangle$,
where $n_{\mathit{fault}}$ is the bug-triggering construct whose execution reaches the code fixed by the historical fix,
$S_{\mathit{key}}$ is the key statement in the regression test that
contains the $n_{\mathit{fault}}$, and
$S_{\mathit{orig}}$ is the complete regression test.
The validation profile is defined as
$V=\langle P_{\mathit{fail}},C_{\mathit{fix}},O_{\mathit{fix}}\rangle$,
where $P_{\mathit{fail}}$ is the failure pattern,
$C_{\mathit{fix}}$ is the fix-derived criterion, expressed as a specification
of expected SQL behavior inferred from the patch and bug description,
and $O_{\mathit{fix}}$ is the fix-derived oracle, specifying a checkable
condition or relation for determining incorrect execution behavior under
the conditions specified in the oracle.
\end{definition}
% \noindent\textbf{Definition~2 (Bug Profile).}
% We define the \emph{Bug Profile} as
% $\Psi=\langle T,V\rangle$, where $T$ is the \emph{trigger profile} and
% $V$ is the \emph{validation profile}. The trigger profile is defined as
% $T=\langle n_{\mathit{fault}},S_{\mathit{key}},S_{\mathit{orig}}\rangle$,
% where $n_{\mathit{fault}}$ is the bug-triggering construct whose execution reaches the code fixed by the historical fix,
% $S_{\mathit{key}}$ is the key statement in the regression test that
% contains the $n_{\mathit{fault}}$, and
% $S_{\mathit{orig}}$ is the complete regression test.
% The validation profile is defined as
% $V=\langle P_{\mathit{fail}},O_{\mathit{fix}},C_{\mathit{fix}}\rangle$,
% where $P_{\mathit{fail}}$ is the failure pattern,
% $O_{\mathit{fix}}$ is the fix-derived oracle, expressed as a specification
% of expected SQL behavior inferred from the patch and bug description,
% and $C_{\mathit{fix}}$ is the fix-derived criterion, specifying a checkable
% condition or relation for determining incorrect execution behavior under
% the stated premises. 

The trigger profile provides the historical context needed for Sibling Test
Generation. $n_{\mathit{fault}}$ identifies the bug-triggering construct
whose syntactic siblings are selected for testing, while $S_{\mathit{key}}$
provides the statement containing that construct and the surrounding SQL
structure needed for mutation. $S_{\mathit{orig}}$ retains the complete
regression test, including the object definitions, data and session settings needed to expose the historical bug. Together, these components allow sibling tests to exercise a different construct while
preserving the bug-triggering conditions. 

The validation profile provides three complementary checks for Sibling Bug Validation. 
$P_{\mathit{fail}}$ supports matching the failure pattern across their executions.
$C_{\mathit{fix}}$ provides the specification inferred from the historical
fix, whose applicability is assessed using the selected sibling's SQL
semantics and execution context. $O_{\mathit{fix}}$ provides a concrete
condition or relation for checking whether the candidate execution is
incorrect under the stated conditions. 
The validation profile provides task-specific context for LLM-based reasoning about sibling bugs and improves the accuracy of sibling-bug detection by reducing false positives caused by valid semantic differences between sibling constructs.
% Although LLMs can draw on broad knowledge of DBMS semantics and open-source implementations, such general knowledge does not provide a problem-specific basis for reasoning about a particular historical bug. Without such a basis, judgments about sibling semantics and correctness may be affected by model uncertainty and lead to unsupported conclusions. 
% \textsc{Reaper} therefore grounds the reasoning in information recovered from the historical fix. 

% Figure~\ref{fig:bug-profiling} illustrates a Bug Profile constructed from the fix for MariaDB bug~\#30324. The regression query \texttt{SELECT COUNT(DISTINCT b) FROM t1 GROUP BY a} exhibits the failure pattern $P_{\mathit{fail}}$: incorrect DISTINCT aggregation over a \texttt{BIT}-typed column. The trigger profile records $n_{\mathit{fault}}=\texttt{COUNT}$ and retains the BIT-valued argument and grouping context associated with the failure. The corresponding patch reveals the correctness requirement that DISTINCT processing preserve the complete logical values of BIT arguments, which the validation profile records as $O_{\mathit{fix}}$. For the historical regression instance, $C_{\mathit{fix}}$ operationalizes this requirement by requiring \texttt{COUNT(DISTINCT b)} to equal the number of distinct logical BIT values in each group. Because sibling constructs may realize different semantics, this historical criterion is not transferred directly to sibling tests. Instead, $O_{\mathit{fix}}$ provides the repair-derived requirement for assessing sibling behavior when the requirement is applicable.

\vspace{1mm}
\noindent\textbf{Bug Profile construction.}
Constructing an accurate Bug Profile from a historical fix is challenging
for two reasons. (i) It is difficult to \textbf{pinpoint the bug-triggering
construct} because the observed failure reflects the interaction of multiple
SQL constructs and their context. In Figure~\ref{fig:bug-profiling}, the
incorrect result involves \texttt{COUNT}, \texttt{DISTINCT},
\texttt{GROUP BY}, and a \texttt{BIT}-typed argument. The description
emphasizes \texttt{BIT} representation, and the patch introduces special
handling for \texttt{FIELD\_TYPE\_BIT}. These clues explain the circumstances
of the failure, but their prominence can lead an LLM to select a triggering
condition rather than the construct corresponding to the repaired handler.
For example, selecting \texttt{DISTINCT} instead of \texttt{COUNT} would
change which sibling handlers are subsequently tested. (ii) It is difficult to \textbf{recover the fix-derived oracle}
because a patch records the concrete repair mechanism, leaving the restored
SQL behavior implicit. In Figure~\ref{fig:bug-profiling}, the patch converts \texttt{BIT}
values to \texttt{BIGINT} in temporary tables, while the failure manifests as
an incorrect result under \texttt{DISTINCT}. The explicit type conversion can
therefore draw an LLM toward implementation details and obscure the expected
behavior: \texttt{DISTINCT} should preserve the complete logical
\texttt{BIT} value.

To address these challenges, \textsc{Reaper} employs three collaborating
LLM agents to construct the Bug Profile. For the first challenge, \textsc{Reaper} adopts a cross-checking mechanism
over independently inferred bug-triggering constructs. The Query Analyzer
infers $n_{\mathit{fault}}^{Q}$ from the original regression test $Q$ and the bug description $D$, while the Patch Analyzer maps the repaired handler in the code patch $\Delta$ to its corresponding SQL construct,
yielding $n_{\mathit{fault}}^{\Delta}$. The Orchestrator confirms
$n_{\mathit{fault}}$ only when the two results refer to the same SQL
construct, thereby reducing the possibility of selecting a surrounding
trigger condition. For the second challenge, \textsc{Reaper} designs the Patch Analyzer
instructions to introduce intermediate reasoning steps between the concrete
code change and the validation profile. These steps guide the derivation of $C_{fix}$ and $O_{fix}$ from
$D$ and $\Delta$.

\noindent\textbf{Query Analyzer.}
The Query Analyzer aims to recover the SQL-side information from the
regression test $Q$ and bug description $D$. To identify
$n_{\mathit{fault}}^{Q}$, \textsc{Reaper} designs instructions (shown in Figure~\ref{fig:bug-profiling}) that direct the LLM to use the diagnostic clues in $D$ to distinguish the bug-triggering
construct in $Q$ from surrounding trigger conditions. The instructions further
require the analyzer to preserve the complete regression test as
$S_{\mathit{orig}}$, extract the complete statement containing
$n_{\mathit{fault}}^{Q}$ as $S_{\mathit{key}}$, and abstract the incorrect
behavior described by $Q$ and $D$ into the failure pattern
$P_{\mathit{fail}}$. In Figure~\ref{fig:bug-profiling}, although
\texttt{COUNT}, \texttt{DISTINCT}, \texttt{GROUP BY}, and the
\texttt{BIT}-typed argument all appear in the failing query, the instructions
direct the analyzer to identify \texttt{COUNT} as
$n_{\mathit{fault}}^{Q}$.

\noindent\textbf{Patch Analyzer.}
The Patch Analyzer aims to recover the repair-side information from the bug
description $D$ and code patch $\Delta$. To obtain both
$n_{\mathit{fault}}^{\Delta}$ and the validation profile, \textsc{Reaper}
designs the instructions with intermediate reasoning steps. Specifically, the
analyzer first traces the repaired handler in $\Delta$ to its corresponding
SQL construct, yielding $n_{\mathit{fault}}^{\Delta}$. It then interprets the
code change together with the information in $D$ and infers
$C_{\mathit{fix}}$ as a specification of the expected SQL behavior. The
instructions also direct the analyzer to identify the conditions and
formulate $O_{\mathit{fix}}$ as a checkable condition or relation for
determining incorrect execution behavior. In
Figure~\ref{fig:bug-profiling}, the modification to
\texttt{Item\_sum\_count\_distinct::setup()} is mapped to \texttt{COUNT}.
By analyzing the \texttt{BIT(N)} representation described in $D$ together
with the \texttt{BIT}-to-\texttt{BIGINT} change in $\Delta$, the analyzer
infers that \texttt{COUNT(DISTINCT ...)} should distinguish
\texttt{BIT}-valued arguments by their logical values.

\newsavebox{\SiblingAlgBox}
\newsavebox{\SiblingCaptionBox}
\newsavebox{\SiblingFigureBox}
\newlength{\SiblingColumnWidth}
\newlength{\SiblingTotalHeight}
\newlength{\SiblingImageHeight}

% Fixed spacing between top/bottom floats and the text
\setlength{\textfloatsep}{10pt}

\begingroup
\captionsetup{hypcap=false}

\begin{figure}[t]
\centering
\setlength{\SiblingColumnWidth}{0.49\linewidth}

% --------------------------------------------------
% Left: typeset and measure the complete algorithm.
% --------------------------------------------------
\begin{lrbox}{\SiblingAlgBox}
\begin{minipage}{\SiblingColumnWidth}
\setlength{\intextsep}{0pt}

\begin{algorithm}[H]
\caption{Sibling Test Generation}
\label{alg:sibling_generation}
\scriptsize
\begin{algorithmic}[1]

\STATE \textbf{input} Trigger Profile
$T=\langle n_{\mathit{fault}},S_{\mathit{key}},S_{\mathit{orig}}\rangle$,
grammar graph $\mathcal{G}_{\delta}$

\STATE \textbf{output} Sibling tests $\mathcal{Q}$

\STATE $\mathit{ast}\gets
\textsc{Parse}(S_{\mathit{key}})$
\label{ln:mutation-parse}

\STATE $\mathit{anchor}\gets
\textsc{MatchConstruct}(\mathit{ast},n_{\mathit{fault}})$
\label{ln:mutation-anchor}

\STATE $A:=\alpha\gets
\textsc{AssociatedProduction}
(\mathit{anchor},\mathcal{G}_{\delta})$
\label{ln:mutation-production}

\STATE $\mathcal{C}\gets
\textsc{CollectSiblingAlternatives}
(A,\mathcal{G}_{\delta},n_{\mathit{fault}})$
\label{ln:mutation-bfs}

\STATE $\mathcal{Q}\gets\varnothing$

\FORALL{$q\in\mathcal{C}$}

    \STATE $\mathit{fragment}\gets
    \textsc{InstantiateAlternative}(q,\mathit{anchor})$
    \label{ln:mutation-fragment}

    \STATE $Q_{\mathit{skel}}\gets
    \textsc{SubstituteOccurrence}
    (S_{\mathit{key}},\mathit{anchor},\mathit{fragment})$
    \label{ln:mutation-substitute}

    \STATE $Q'\gets
    \textsc{Concretize}
    (Q_{\mathit{skel}},S_{\mathit{orig}},\mathcal{G}_{\delta})$
    \label{ln:mutation-concretize}

    \IF{$Q'\neq\bot$}
        \STATE $\mathcal{Q}\gets\mathcal{Q}\cup\{Q'\}$
        \label{ln:mutation-collect}
    \ENDIF

\ENDFOR

\STATE \textbf{return} $\mathcal{Q}$

\end{algorithmic}
\end{algorithm}
\end{minipage}
\end{lrbox}

% --------------------------------------------------
% Measure the figure caption at its final width.
% --------------------------------------------------
\begin{lrbox}{\SiblingCaptionBox}
\begin{minipage}{\SiblingColumnWidth}
\caption{Sibling Test Generation.}
\label{fig:sibling-generation}
\end{minipage}
\end{lrbox}

% Total height of the algorithm, including its caption.
\setlength{\SiblingTotalHeight}{%
  \dimexpr
  \ht\SiblingAlgBox+\dp\SiblingAlgBox
  \relax
}

% Available image height after accounting for the caption.
\setlength{\SiblingImageHeight}{%
  \dimexpr
  \SiblingTotalHeight
  -\ht\SiblingCaptionBox-\dp\SiblingCaptionBox
  \relax
}

% --------------------------------------------------
% Right: construct a column with exactly the same height.
% --------------------------------------------------
\setbox\SiblingFigureBox=\vbox to
  \dimexpr\SiblingTotalHeight-\dp\SiblingCaptionBox\relax{%
    \offinterlineskip
    \hbox to \SiblingColumnWidth{%
      \hfil
      \includegraphics[
        width=\SiblingColumnWidth,
        height=\SiblingImageHeight,
        keepaspectratio
      ]{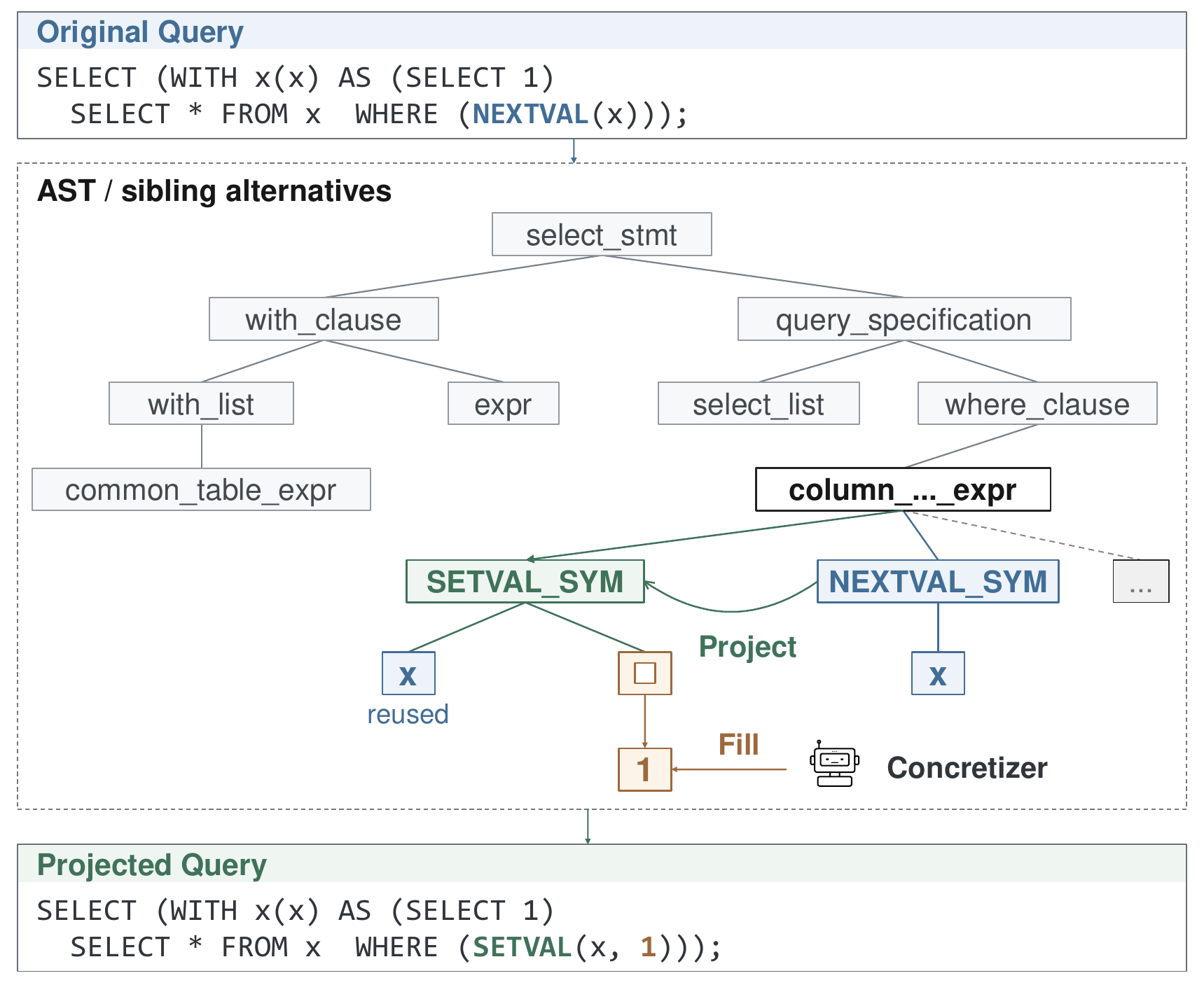}%
      \hfil
    }%
    \vfil
    \hbox{\usebox{\SiblingCaptionBox}}%
  }

% Align both top edges; equal total heights align the bottoms.
\noindent
\raisebox{-\height}{\usebox{\SiblingAlgBox}}%
\hfill
\raisebox{-\height}{\usebox{\SiblingFigureBox}}%

\Description{The sibling test generation algorithm alongside an
example that replaces NEXTVAL with SETVAL, reuses the argument x,
and fills the additional operand with 1.}

\par

\end{figure}
\endgroup

\noindent\textbf{Orchestrator.}
The Orchestrator coordinates the two analyzers through the cross-checking
mechanism. Specifically, it compares $n_{\mathit{fault}}^{Q}$ and
$n_{\mathit{fault}}^{\Delta}$ together with their corresponding findings and
confirms $n_{\mathit{fault}}$ when the two analyses refer to the same SQL
construct. When they disagree, the Orchestrator returns the conflicting
findings to the analyzers for re-analysis and filters out fixes whose
disagreement remains unresolved. After confirming $n_{\mathit{fault}}$, it
constructs the trigger profile from $S_{\mathit{orig}}$,
$S_{\mathit{key}}$, and $n_{\mathit{fault}}$, and the validation profile
from $P_{\mathit{fail}}$, $C_{\mathit{fix}}$, and $O_{\mathit{fix}}$.
Together, the two profiles constitute the final Bug Profile $\Psi$.

\subsection{Sibling Test Generation}
\label{sec:mutation}

With the trigger profile recovered from a historical fix, \textsc{Reaper}
extends testing from the reported construct to its syntactic siblings while
retaining the context that makes the historical bug relevant. The challenge
is to redirect execution to a sibling handler while preserving the
bug-triggering conditions inherited from the historical test. To achieve
this, \textsc{Reaper} applies \emph{Grammar-Guided Sibling Mutation} to
identify grammar-defined siblings and construct statement skeletons through
localized substitution. Guided by the trigger profile, an LLM-based
\emph{Concretizer} then performs \emph{Trigger-Constrained Concretization}
to complete these skeletons with sibling-specific operands, bindings, and
execution requirements while preserving the historical bug-triggering
context.

\subsubsection{Grammar-Guided Sibling Mutation}
\textsc{Reaper} uses the syntactic-sibling relation defined in
\S\ref{sec:backend} to identify alternatives to the bug-triggering
construct $n_{\mathit{fault}}$ recorded in the trigger profile.
Algorithm~\ref{alg:sibling_generation} summarizes the complete sibling
test generation procedure. To support sibling discovery, \textsc{Reaper}
represents the target DBMS grammar as a grammar graph
$\mathcal{G}_{\delta}$, where production alternatives are connected
through the nonterminals they reference. Given the trigger profile $T$ and $\mathcal{G}_{\delta}$, \textsc{Reaper} locates the
occurrence of $n_{\mathit{fault}}$ associated with the historical failure
in the AST of $S_{\mathit{key}}$ and maps it to the corresponding grammar
production $A := \alpha$. 
Starting from the alternative productions of $A$, \textsc{Reaper}
performs a breadth-first search (BFS) that expands unit productions
but does not descend into the internal nonterminals of concrete
constructs (Line~\ref{ln:mutation-bfs}). A concrete alternative is
retained as a sibling candidate if its construct differs from
$n_{\mathit{fault}}$. Each production is processed at most once,
and the search terminates when the queue is empty.

For each sibling alternative, \textsc{Reaper} uses its production as
the structural template for a replacement fragment.
\textsc{InstantiateAlternative} matches nonterminal occurrences by
symbol and occurrence order, filling matched occurrences with the
corresponding historical subtrees unchanged. Unmatched nonterminal
occurrences remain as placeholders for sibling-specific components.
The fragment then replaces only the selected historical occurrence
in $S_{\mathit{key}}$ to obtain the statement skeleton
$Q_{\mathit{skel}}$ (Line~\ref{ln:mutation-substitute}). This local
replacement preserves both the historical subtrees reused within the
fragment and the surrounding statement structure outside it. The
complete regression test $S_{\mathit{orig}}$ is passed separately to
the Concretizer, which uses the historical setup and execution context
to complete the skeleton and satisfy sibling-specific semantic
requirements.

Figure~\ref{fig:sibling-generation} illustrates this process for
\texttt{NEXTVAL}. The grammar identifies \texttt{SETVAL} as a sibling
construct. Since both alternatives contain \texttt{table\_ident},
\textsc{Reaper} reuses the historical subtree \texttt{x}. The additional
value expression required by \texttt{SETVAL} has no historical
counterpart and is left as a placeholder, yielding
\texttt{SETVAL(x,~$\square$)}. Replacing the selected \texttt{NEXTVAL(x)} occurrence with this fragment produces the sibling statement skeleton while preserving the historical bug-triggering conditions.

\subsubsection{Trigger-Constrained Concretization}

\textsc{Reaper} employs an LLM-based \emph{Concretizer} to complete
the sibling statement skeleton $Q_{\mathit{skel}}$
(Line~\ref{ln:mutation-concretize}). The Concretizer receives the
skeleton constructed from $n_{\mathit{fault}}$ and
$S_{\mathit{key}}$, together with the complete historical regression
test $S_{\mathit{orig}}$ from the trigger profile. The skeleton
specifies the selected sibling and its unresolved components, while
$S_{\mathit{orig}}$ provides the historical statements, definitions,
and bindings required to complete the sibling test without discarding
the historical bug-triggering context.

The Concretizer first compares $Q_{\mathit{skel}}$ with
$S_{\mathit{key}}$ to identify the operands and additional requirements
introduced by the sibling construct. For unresolved components, it consults
the grammar graph $\mathcal{G}_{\delta}$ to restrict their completion to
alternatives permitted by the corresponding grammar productions. It then
examines $S_{\mathit{orig}}$ to determine which definitions and bindings are
used by the reused subtrees and must therefore be retained. Based on this
information, the Concretizer completes the unresolved components while
introducing new elements only where required by the sibling construct.
A completion is accepted only if it conforms to the grammar and satisfies
the additional requirements while retaining the statements and bindings
relevant to the historical bug-triggering context; otherwise, the
Concretizer returns $\bot$ and the candidate is discarded.

Figure~\ref{fig:sibling-generation} illustrates this process for
\texttt{SETVAL(x,~$\square$)}. The Concretizer first identifies the
additional value argument required by \texttt{SETVAL}. It then examines
$S_{\mathit{orig}}$ and identifies that \texttt{x} refers to a CTE whose
pure-alias binding is required to trigger the historical bug. Replacing
\texttt{x} with an ordinary sequence name would satisfy the object
requirement of \texttt{SETVAL}, but would remove this historical
binding. The Concretizer therefore retains the CTE definition and its
reference and supplies \texttt{1} for the unresolved value expression,
yielding \texttt{SETVAL(x,~1)}.

\begin{figure}[t]
    \centering
    \includegraphics[width=\textwidth]{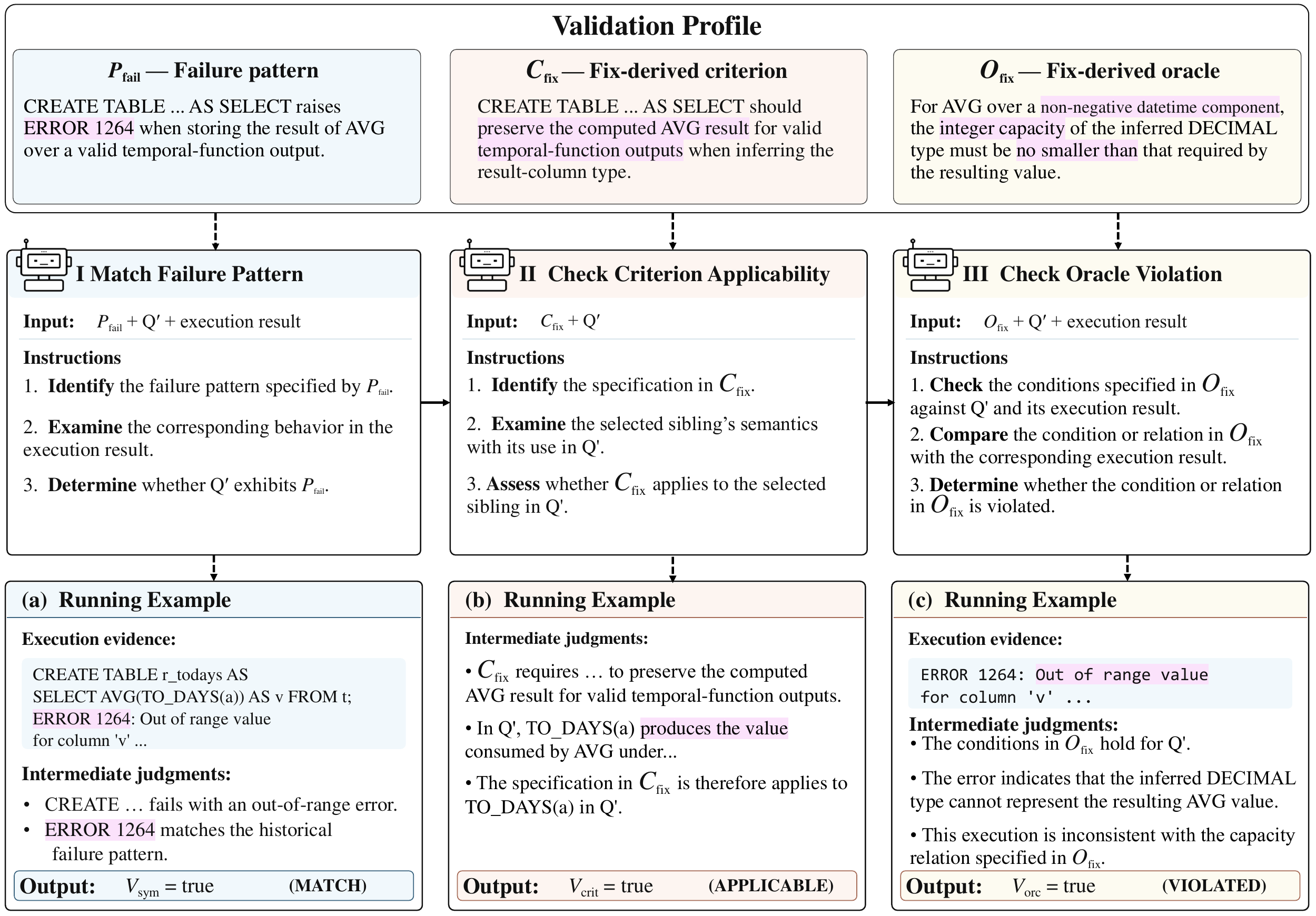}
    \caption{LLM-based Sibling Bug Validation guided by the validation profile.}
    \label{fig:sibling-validation}
    \vspace{-3mm}
\end{figure}

\subsection{Sibling Bug Validation}
\label{sec:oracle}

Syntactic siblings may implement different semantics and consequently produce
different valid results. Sibling bug validation therefore requires semantic
reasoning about the execution result of a sibling test $Q'$, beyond direct
comparison with the historical execution result. LLMs can support this
reasoning using their knowledge of SQL semantics and open-source DBMS
implementations. However, such general knowledge alone may be insufficient
for judging a particular sibling test. Without information specific to the
historical bug, the model may rely on unsupported assumptions about the
selected sibling and its execution result. 

To provide a basis for these judgments, \textsc{Reaper} uses the
Validation Profile
$V=\langle P_{\mathit{fail}},C_{\mathit{fix}},O_{\mathit{fix}}\rangle$
recovered from the historical fix to guide three LLM agents. Each agent
follows intermediate reasoning steps organized around the corresponding
profile component. The three judgments serve distinct purposes. The first
uses $P_{\mathit{fail}}$ to restrict validation to executions exhibiting
the historical failure pattern. The second uses $C_{\mathit{fix}}$ to
assess whether the specification recovered from the historical fix applies
to the selected sibling. The third evaluates $O_{\mathit{fix}}$ to
establish whether the execution result is incorrect under the conditions
specified in the oracle, since criterion applicability alone does not
establish incorrect execution behavior. Figure~\ref{fig:sibling-validation} uses MariaDB MDEV-33496 to illustrate
the validation process. The historical bug occurs when
\texttt{CREATE TABLE ... AS SELECT} derives a result column for
\texttt{AVG(YEAR(a))}. The implementation did not correctly account for
the range of non-negative datetime-component functions when deriving the
\texttt{DECIMAL} type for \texttt{AVG}, producing a column with
insufficient integer capacity and raising \texttt{ERROR 1264} for
otherwise valid inputs. The historical fix corrects the type handling
for these functions so that the derived \texttt{DECIMAL} type can
represent the resulting value. Starting from this fix, \textsc{Reaper}
generates a sibling test by replacing \texttt{YEAR(a)} with
\texttt{TO\_DAYS(a)} and validates its execution using the recovered
validation profile.

\noindent\textbf{Stage I: Match Failure Pattern.}
The first agent determines whether the execution result of $Q'$ exhibits
the historical failure pattern, restricting further validation to candidates
relevant to the historical bug. Given $P_{\mathit{fail}}$, $Q'$, and its
execution result, the prompt guides the agent through three reasoning
steps. First, the agent identifies the failure pattern specified by
$P_{\mathit{fail}}$. Second, it examines the corresponding behavior in
the execution result. Finally, it determines whether $Q'$ exhibits
$P_{\mathit{fail}}$ based on these intermediate judgments. The comparison
considers the failure described by the pattern and the operation in which
it occurs, without requiring identical concrete outputs. In
Figure~\ref{fig:sibling-validation}(a), $P_{\mathit{fail}}$ describes
\texttt{ERROR 1264} when
\texttt{CREATE TABLE ... AS SELECT} stores an \texttt{AVG} result.
The agent identifies an out-of-range error in the execution of the
statement containing \texttt{AVG(TO\_DAYS(a))} and relates this error
to the historical failure pattern. It therefore returns \textsc{Match},
for which $V_{\mathit{sym}}(Q')$ is true. A \textsc{No\_Match}
decision excludes the candidate from further validation.

\vspace{1mm}
\noindent\textbf{Stage II: Check Criterion Applicability.}
The second agent determines whether the specification in
$C_{\mathit{fix}}$ applies to the selected sibling. This judgment is
necessary because a matching failure pattern does not establish that the fix-derived criterion is applicable to the sibling's semantics. Given
$C_{\mathit{fix}}$ and $Q'$, the prompt first directs the agent to
identify the specification in $C_{\mathit{fix}}$. Second, the agent
examines the selected sibling's semantics together with its use in $Q'$.
Finally, it assesses whether the specification applies to the selected
sibling in that statement. In
Figure~\ref{fig:sibling-validation}(b), $C_{\mathit{fix}}$ specifies that
\texttt{CREATE TABLE ... AS SELECT} should preserve the computed
\texttt{AVG} result for valid temporal-function outputs when inferring
the result-column type. The agent identifies that \texttt{TO\_DAYS(a)}
produces the values consumed by \texttt{AVG} within
\texttt{CREATE TABLE ... AS SELECT}. It therefore concludes that the
specification applies to this use of \texttt{TO\_DAYS(a)} and returns
\textsc{Applicable}, for which $V_{\mathit{crit}}(Q')$ is true.
A \textsc{Not\_Applicable} decision excludes the candidate from the
next stage.

\vspace{1mm}
\noindent\textbf{Stage III: Check Oracle Violation.}
The third agent determines whether the execution result violates the
condition or relation specified in $O_{\mathit{fix}}$. Given
$O_{\mathit{fix}}$, $Q'$, and its execution result, the prompt organizes
the judgment into three reasoning steps. First, the agent checks the
conditions specified in $O_{\mathit{fix}}$ against $Q'$ and its execution
result. Second, it compares the condition or relation in
$O_{\mathit{fix}}$ with the corresponding execution result. Finally, it
assesses the oracle result based on the condition checks and evaluation.
In Figure~\ref{fig:sibling-validation}(c), $O_{\mathit{fix}}$ requires
that, for \texttt{AVG} over a non-negative datetime component, the
integer capacity of the inferred \texttt{DECIMAL} type be no smaller
than that required by the resulting value. In the illustrated reasoning,
the agent first determines that the conditions in $O_{\mathit{fix}}$
hold for $Q'$. It then interprets the out-of-range error as indicating
that the inferred \texttt{DECIMAL} type cannot represent the resulting
\texttt{AVG} value. On this basis, the agent concludes that the inferred
integer capacity is smaller than that required by the result and returns
\textsc{Violated}. Accordingly, $V_{\mathit{orc}}(Q')$ is true only
when the conditions hold and the agent determines that the condition
or relation in $O_{\mathit{fix}}$ is violated.

\textsc{Reaper} reports a sibling bug only when the three agents return
\textsc{Match}, \textsc{Applicable}, and \textsc{Violated}, respectively:
\begin{equation}
\operatorname{Report}(Q')
\Leftrightarrow
V_{\mathit{sym}}(Q')
\land
V_{\mathit{crit}}(Q')
\land
V_{\mathit{orc}}(Q').
\label{eq:sibling-validation}
\end{equation}

\begin{table}[t]
\caption{Unknown bugs found by \textsc{Reaper}.}
\label{tab:bug_summary}
\centering
\small
\setlength{\tabcolsep}{5pt}

\begin{tabular}{lrrrrrrr}
\toprule
\textbf{Database System} &
\multicolumn{3}{c}{\textbf{Crash/Fatal Bugs}} &
\multicolumn{3}{c}{\textbf{Logic Bugs}} &
\textbf{Total} \\

\cmidrule(lr){2-4}
\cmidrule(lr){5-7}

&
\textbf{Pending} &
\textbf{Confirmed} &
\textbf{Fixed} &
\textbf{Pending} &
\textbf{Confirmed} &
\textbf{Fixed} &
\\

\midrule

MariaDB
& 0 & 4 & 1
& 0 & 19 & 2
& 23 \\

MySQL
& 0 & 0 & 0
& 0 & 15 & 1
& 15 \\

OceanBase
& 0 & 2 & 0
& 3 & 4 & 0
& 9 \\

CockroachDB
& 0 & 5 & 1
& 1 & 8 & 8
& 14 \\

\midrule

\textbf{Total}
& \textbf{0}
& \textbf{11}
& \textbf{2}
& \textbf{4}
& \textbf{46}
& \textbf{11}
& \textbf{61} \\

\bottomrule
\end{tabular}
\vspace{-0.5em}
\end{table}

\section{Evaluation}
\label{4-experiment}

To evaluate the effectiveness of \textsc{Reaper} in detecting sibling bugs in real-world DBMSs, we address the following four research questions:

\begin{itemize}
\item \textbf{RQ1:} How effective is \textsc{Reaper} at discovering sibling bugs in real-world DBMSs?
\item \textbf{RQ2:} How does \textsc{Reaper} compare with existing DBMS fuzzers?
\item \textbf{RQ3:} How do \textsc{Reaper}'s key test-generation components contribute to sibling-bug discovery?
\item \textbf{RQ4:} How effectively does \textsc{Reaper} distinguish logic sibling bugs from benign behavior?
\end{itemize}

\subsection{Evaluation Setup}
\label{sec:eval-setup}

\subsubsection{Environment.}
We conducted all experiments on a machine running 64-bit Ubuntu 22.04.5 LTS, equipped with an Intel(R) Xeon(R) Gold 6454S CPU with 128 cores and 512 GB of RAM. All LLM-based components in \textsc{Reaper} use DeepSeek-V4-Flash through its official API.

\subsubsection{Target Databases.}
We evaluate \textsc{Reaper} on four production-grade relational DBMSs: MySQL (v9.6.0), MariaDB (v13.2), OceanBase (v4.5.1), and CockroachDB (v26.3.0). The selected systems cover both centralized and distributed architectures and differ in their SQL grammar and execution implementations, allowing us to evaluate \textsc{Reaper} across heterogeneous DBMS designs. The coverage comparison is conducted on MySQL, MariaDB, and OceanBase. CockroachDB is excluded because its branch coverage could not be measured under the same instrumentation setup.

\subsubsection{Baselines.}
We compare \textsc{Reaper} against three state-of-the-art DBMS testing approaches: SQLancer~\cite{pqs,norec,tlp}, Squirrel~\cite{squirrel}, and Argus~\cite{argus}. These baselines represent complementary strategies for DBMS testing, including generation-based testing, structure-aware mutation, and LLM-assisted oracle discovery and test generation. SQLancer generates queries from scratch and detects logic bugs using DBMS testing oracles such as TLP~\cite{tlp} and NoREC~\cite{norec}. In our experiments, we use TLP for MySQL and NoREC for the remaining DBMS targets. Squirrel is a mutation-based fuzzer that translates seed queries into an intermediate representation and applies structure-aware mutations to explore new execution paths. Argus uses LLMs to propose pairs of equivalent Constrained Abstract Queries (CAQs), verifies their equivalence using a SQL equivalence solver, and instantiates the verified pairs with reusable SQL snippets to generate concrete test cases~\cite{argus}.

\subsubsection{Historical Bug Corpus.}
To construct the historical bug corpus, we mine bug-fixing commits from the GitHub repositories of the four target DBMSs and retain commits for which the artifacts required by Bug Profile Construction can be recovered. For each retained commit, we extract the associated description, regression test, and code patch. In total, this collection stage yields 4{,}487 historical bug-fixing commits: 1{,}713 from MySQL, 1{,}139 from OceanBase, 1{,}112 from MariaDB, and 523 from CockroachDB. These commits form the input corpus for Bug Profile Construction.

\begin{figure}[t]
    \centering
    \includegraphics[width=\textwidth]{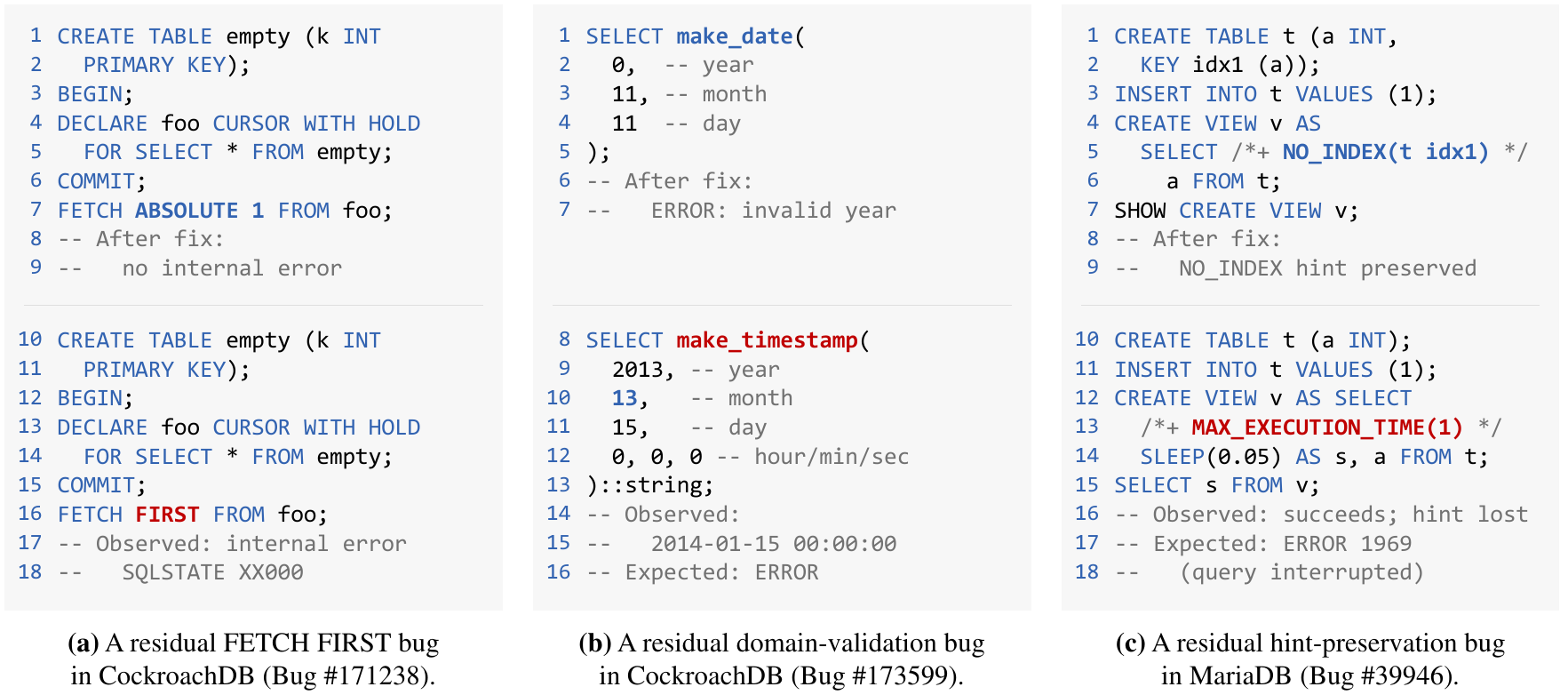}
    \caption{Representative sibling bugs discovered by \textsc{Reaper}.
Each column compares a repaired historical query with its sibling test
that exposes a residual bug.}
    \label{fig:bug_diversity_examples}
    \vspace{-2mm}
\end{figure}

\vspace{-2mm}
\subsection{RQ1: Effectiveness in Sibling-Bug Discovery}
\label{sec:rq1}

\noindent\textbf{Overall results.}
\textsc{Reaper} discovered 61 previously unknown sibling bugs across the four
evaluated DBMSs: 23 in MariaDB, 15 in MySQL, 9 in OceanBase, and 14 in
CockroachDB. As summarized in Table~\ref{tab:bug_summary}, 57 bugs have been
confirmed by upstream developers, of which 13 have already been fixed, while
the remaining 4 are still under analysis. Notably, a single historical
fix can expose residual bugs in multiple sibling handlers. For example,
starting from the historical fix for \texttt{NEXTVAL}, \textsc{Reaper}
exposed the same residual issue in both \texttt{LASTVAL} and \texttt{SETVAL}.
We conservatively consolidate such manifestations when they correspond to the
same underlying bug, so they are counted as one sibling bug rather than
multiple discoveries. Hence, the 61 bugs represent distinct reported bugs,
while the affected sibling handlers and triggering tests are more numerous,
demonstrating that \textsc{Reaper} can extend a localized historical fix to
systematically uncover unpatched behavior across its sibling handlers. Beyond the reported bugs, \textsc{Reaper} extends each historical regression
test from the repaired construct to its grammar-defined sibling set. Under the
recovered bug-triggering context, the generated sibling tests enable systematic
validation of the related sibling handlers for the same historical bug. This
turns a regression test for a single repaired construct into a sibling-level
regression task, allowing the original test and its generated sibling tests to
be executed together in existing regression-testing and CI workflows.
% These bugs span multiple DBMS components, including the parser, optimizer,
% executor, and built-in routines, demonstrating that sibling-handler
% inconsistencies are not confined to a particular implementation layer.
% Across these cases, the original regression tests successfully exercise the
% repaired paths, while related sibling handlers continue to exhibit behavior
% inconsistent with the repaired requirement.

\begin{figure}[t]
\centering
\begin{tikzpicture}[
  x=0.55cm, y=0.55cm,
  num/.style={font=\small\bfseries, text=black, inner sep=0pt},
  zero/.style={font=\scriptsize, text=black, inner sep=0pt},
  leg/.style={font=\scriptsize, anchor=west, inner sep=0pt}
]

\definecolor{vennBlue}{HTML}{9ECAE1}
\definecolor{vennBlueLine}{HTML}{2171B5}
\definecolor{vennOrange}{HTML}{FDD0A2}
\definecolor{vennOrangeLine}{HTML}{D94801}
\definecolor{vennGreen}{HTML}{A1DAB4}
\definecolor{vennGreenLine}{HTML}{238B45}
\definecolor{vennPink}{HTML}{F2B6D8}
\definecolor{vennPinkLine}{HTML}{AE3A94}

\def\rx{3.0}
\def\ry{1.55}

\begin{scope}[shift={(0.8,0)}]

\filldraw[
  fill=vennBlue,
  draw=vennBlueLine,
  fill opacity=0.32,
  line width=0.7pt,
  rotate around={45:(-0.95,0.35)}
]
(-0.95,0.35) ellipse [x radius=\rx, y radius=\ry];

\filldraw[
  fill=vennOrange,
  draw=vennOrangeLine,
  fill opacity=0.32,
  line width=0.7pt,
  rotate around={45:(-0.38,-0.05)}
]
(-0.38,-0.05) ellipse [x radius=\rx, y radius=\ry];

\filldraw[
  fill=vennGreen,
  draw=vennGreenLine,
  fill opacity=0.32,
  line width=0.7pt,
  rotate around={-45:(0.38,-0.05)}
]
(0.38,-0.05) ellipse [x radius=\rx, y radius=\ry];

\filldraw[
  fill=vennPink,
  draw=vennPinkLine,
  fill opacity=0.32,
  line width=0.7pt,
  rotate around={-45:(0.95,0.35)}
]
(0.95,0.35) ellipse [x radius=\rx, y radius=\ry];

\node[num] at (-2.62, 0.10) {38};
\node[num] at ( 2.62, 0.10) {7};
\node[zero] at (-0.80,-1.86) {0};
\node[num] at ( 0.80,-1.86) {1};

\node[zero] at (-1.88,-1.10) {0};
\node[zero] at ( 1.88,-1.10) {0};
\node[zero] at (-1.70, 1.28) {0};
\node[zero] at ( 1.70, 1.28) {0};
\node[zero] at ( 0.00, 2.39) {0};
\node[zero] at ( 0.00,-1.32) {0};
\node[zero] at (-1.03,-0.20) {0};
\node[zero] at ( 1.03,-0.20) {0};
\node[zero] at (-0.94, 1.92) {0};
\node[zero] at ( 0.94, 1.92) {0};
\node[zero] at ( 0.00, 0.82) {0};

\end{scope}

\begin{scope}[shift={(4.25,-1.00)}]

\draw[
  rounded corners=2pt,
  draw=black!30,
  fill=white,
  fill opacity=0.92,
  line width=0.4pt
]
(0,1.85) rectangle (3.95,3.95);

\def\sw{0.34}

\foreach \i/\fc/\lc/\t in {
  0/vennBlue/vennBlueLine/{Reaper (38)},
  1/vennPink/vennPinkLine/{SQLancer (7)},
  2/vennOrange/vennOrangeLine/{Squirrel (0)},
  3/vennGreen/vennGreenLine/{Argus (1)}
}{
  \pgfmathsetmacro\yy{3.50-\i*0.48}
  \draw[
    rounded corners=1pt,
    fill=\fc,
    draw=\lc,
    fill opacity=0.55,
    line width=0.4pt
  ]
  (0.20,\yy) rectangle +(\sw,\sw);

  \node[leg] at (0.66,\yy+\sw/2) {\t};
}

\end{scope}

\end{tikzpicture}
\vspace{-0.8em}
\caption{The distribution of unique bugs detected by \textsc{Reaper} and the baselines.}
\vspace{-3mm}
\label{fig:venn-baselines}
\end{figure}
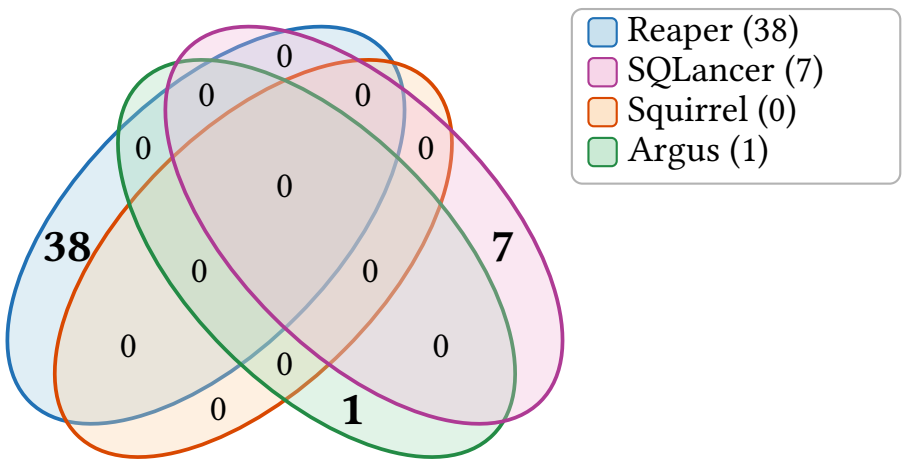

\noindent\textbf{Bug Significance and Impact.}
The reported bugs span silently incorrect results, crashes, and fatal internal
errors. Their upstream severity assessments and subsequent fixes further
indicate their practical impact. All 15 MySQL bugs listed in
Table~\ref{tab:bug_summary} carry Oracle's S3 severity rating, while most
confirmed MariaDB bugs were rated \emph{major}. Thirteen bugs have already
been patched upstream.

MariaDB Bug~\#39063 provides a representative example of a sibling bug.
The historical bug affected \texttt{NEXTVAL}, where the corresponding repair
introduced a \texttt{table\_list->is\_pure\_alias()} check. \textsc{Reaper}
subsequently exposed the same missing handling in the \texttt{LASTVAL} and
\texttt{SETVAL} sibling handlers. MariaDB upgraded the report from
\emph{major} to \emph{critical}, and the developer explicitly referred to
the historical \texttt{NEXTVAL} repair when suggesting the fix: ``try adding the
\texttt{table\_list->is\_pure\_alias()} block to
\texttt{Item\_func\_setval::val\_int}, same as
\texttt{Item\_func\_nextval::val\_int}? Add a test case appended to where the
previous fix was and create a pull request.''

Cross-version reproduction further indicates that some sibling bugs remain
present across multiple maintained releases. For Bug~\#39598, a MariaDB
developer reproduced the issue on four versions: ``I could reproduce the issue with testcase provided in report on mariadb 10.11, 11.4, 11.8 and 12.3\ldots{} This inconsistency
occurs in the error path and can break tools or scripts that expect a uniform
format in the Table column. Attached the mtr test to repro.'' Notably, MariaDB 10.11 became generally available in February 2023, showing that the issue remained present across release lines spanning more than three years.

Several reports also received rapid upstream responses. For CockroachDB
Bug~\#171234, the developer took ownership of the report on the day of
disclosure and identified the missing guard in both
\texttt{width\_bucket} overloads: ``Taking this up. The guard is missing from both the numeric and int overloads
of \texttt{width\_bucket} in
\texttt{pkg/sql/sem/builtins/math\_builtins.go}. Will open a draft PR
shortly.'' An upstream fix was submitted on the following day.

\vspace{1mm}
\noindent\textbf{Bug Diversity.}
The 61 sibling bugs cover a broad range of failure manifestations and DBMS
components. Among them, 50 are logic bugs that complete execution but
manifest in diverse forms of incorrect behavior, while the remaining 11
result in crashes or fatal internal errors. These bugs arise across multiple
stages of DBMS processing, including query analysis, optimization,
temporary-table processing, execution, and built-in routines. To better
illustrate the diversity of these sibling bugs, we present three
representative cases showcasing the different implementation locations and
validation conditions encountered by \textsc{Reaper}. Specifically, the
selected cases span cursor handling, datetime construction, and hint
preservation across different DBMS components and failure manifestations.

\noindent\emph{Implementation-location diversity.}
The three cases show how a historical regression test can leave related
handlers untested across different implementation structures. In
Figure~\ref{fig:bug_diversity_examples}(a), \texttt{FETCH ABSOLUTE}
and \texttt{FETCH FIRST} reach separate cursor-dispatch branches.
Discovering the residual bug requires changing the cursor operation while
retaining the empty-table, \texttt{WITH HOLD}, and cross-transaction
conditions. Grammar-Guided Sibling Mutation performs this localized
substitution, directing the historical trigger to the sibling branch that
lacks the check. In
Figure~\ref{fig:bug_diversity_examples}(b), the missing validation occurs
in a different built-in routine: \texttt{make\_timestamp} accepts an
out-of-domain calendar component even though the historical fix corrected
such handling in \texttt{make\_date}. Here, sibling selection identifies
the target routine, while Trigger-Constrained Concretization supplies its
additional time arguments without removing the bug-triggering condition.
In Figure~\ref{fig:bug_diversity_examples}(c), the two hints follow
different handling paths during view creation. The Concretizer supplies
the time limit and delay needed to exercise
\texttt{MAX\_EXECUTION\_TIME}, while retaining the historical
view-creation context. Across these cases, the syntactic-sibling relation
determines which alternative handler to test, and the trigger profile
constrains the sibling tests to retain the conditions relevant to the bug.

\noindent\emph{Validation-condition diversity.}
The contrast among the three cases illustrates why direct differential
comparison with the historical execution result is insufficient for
validating sibling bugs. In Figure~\ref{fig:bug_diversity_examples}(a),
the sibling exposes an internal error under the preserved bug-triggering
conditions, allowing Stage~I to associate the execution with the
historical bug through $P_{\mathit{fail}}$. The other two cases are more
subtle because the sibling executions complete successfully. In
Figure~\ref{fig:bug_diversity_examples}(b), the concrete input changes
from \texttt{year=0} to \texttt{month=13}, and
\texttt{make\_timestamp} naturally produces a different form of result
from \texttt{make\_date}. Comparing their concrete outputs would
therefore provide no useful oracle. Instead, Stage~II establishes that
$C_{\mathit{fix}}$ also applies to the sibling, while Stage~III uses
$O_{\mathit{fix}}$ to determine that returning a normalized timestamp
for an out-of-domain calendar component is incorrect. A similar issue
appears in Figure~\ref{fig:bug_diversity_examples}(c):
\texttt{NO\_INDEX} and \texttt{MAX\_EXECUTION\_TIME} have different
semantics, so their execution results are not expected to be identical.
Reaper instead checks whether $C_{\mathit{fix}}$ applies to the sibling
hint and then uses $O_{\mathit{fix}}$ to determine whether the hint is
preserved and takes effect. These cases show that the staged validation
does not require a sibling test to reproduce the historical execution
result; it uses the Validation Profile to determine whether the sibling
execution represents the same bug under its own semantics.

\vspace{-2mm}
\subsection{RQ2: Baseline Comparison}
\label{sec:rq2}
To assess the effectiveness of \textsc{Reaper}, we compare it with three
state-of-the-art DBMS testing methods: \textsc{SQLancer}, \textsc{Sq{}uirrel},
and \textsc{Argus}. These baselines represent complementary testing paradigms,
including query generation, mutation-based exploration, and LLM-guided test
synthesis. Each tool runs for 8 hours per target DBMS on identical
hardware, using the same seeds when required and its default configuration
without additional tuning.
We measure bug-finding effectiveness by the number of unique bugs discovered
in the controlled comparison. We additionally record branch coverage as well
as syntactic and semantic validity to characterize the explored execution
space.

\begin{figure}[t]
\centering
\subfloat[MySQL]{
  \includegraphics[width=0.29\linewidth]{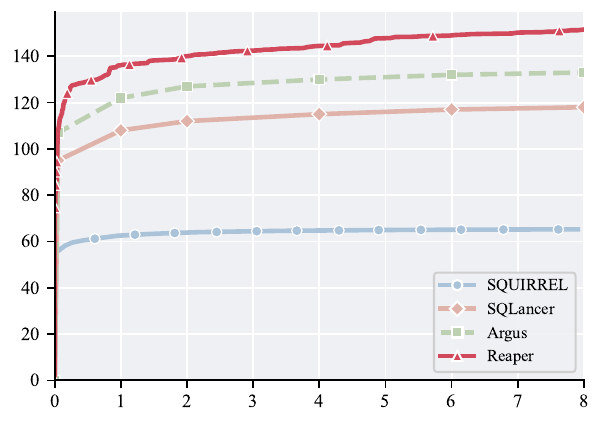}
  \label{fig:cov-mysql}
}
\hfil
\subfloat[MariaDB]{
  \includegraphics[width=0.29\linewidth]{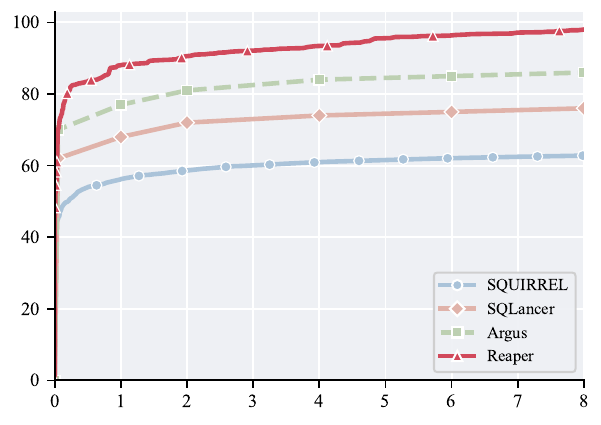}
  \label{fig:cov-mariadb}
}
\hfil
\subfloat[OceanBase]{
  \includegraphics[width=0.29\linewidth]{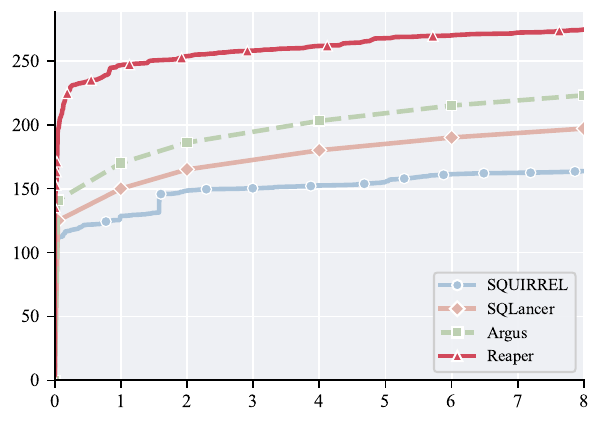}
  \label{fig:cov-oceanbase}
}
\vspace{-0.8em}
\caption{Code branch coverage (k) on MySQL, MariaDB, and OceanBase.}
\label{fig:coverage-over-time}
\vspace{-0.8em}
\end{figure}

\vspace{1mm}
\noindent\textbf{Bug-Finding Capability.}
Figure~\ref{fig:venn-baselines} summarizes the unique bugs discovered
in the controlled comparison. \textsc{Reaper} discovers 38 bugs, compared with
7 for \textsc{SQLancer}, 1 for \textsc{Argus}, and none for \textsc{Squirrel},
with no overlap between \textsc{Reaper} and the baselines. The key distinction lies in the defect space targeted by each approach. Existing approaches explore a general SQL input space through generation or mutation, whereas \textsc{Reaper} replaces the historical bug-triggering construct with its syntactic siblings while preserving the associated bug-triggering context.
Consequently, its search is concentrated on handlers that are structurally
related to a repaired path but may not have received the same fix. This Bug-Profile-guided sibling testing exposes residual inconsistencies that general-purpose generation and mutation do not explicitly target.

\vspace{1mm}
\noindent\textbf{Code Coverage.}
Figure~\ref{fig:coverage-over-time} shows branch-coverage growth on MySQL,
MariaDB, and OceanBase over the 8-hour campaigns.
We exclude CockroachDB due to the instrumentation limitation discussed in
Section~\ref{sec:eval-setup}.
\textsc{Reaper} achieves higher branch coverage than all baselines on all three systems throughout
each campaign and finishes at approximately 151K, 98K, and 274K branches on
MySQL, MariaDB, and OceanBase, respectively.

By comparison, the strongest baseline \textsc{Argus} reaches 133K, 86K, and
222K; \textsc{SQLancer} reaches 118K, 76K, and 197K; and
\textsc{Squirrel} reaches 65K, 63K, and 165K. Relative to the strongest
baseline on each system, these final coverage values correspond to gains of
13.5\%, 14.0\%, and 23.4\%. The additional coverage reached by \textsc{Reaper} reflects its ability to traverse implementation paths associated with repaired constructs and their syntactic siblings. These paths are not merely uncovered SQL features;
they represent alternative handlers that may share the repaired requirement
but evolve inconsistently during maintenance.

\begin{figure}[t]
\centering
\subfloat[Syntactic validity]{
  \includegraphics[width=0.38\linewidth]{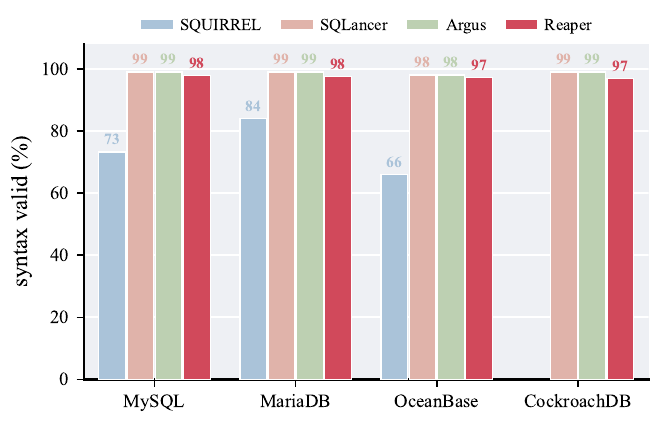}
  \label{fig:syn-valid}
}
\hfil
\subfloat[Semantic validity]{
  \includegraphics[width=0.38\linewidth]{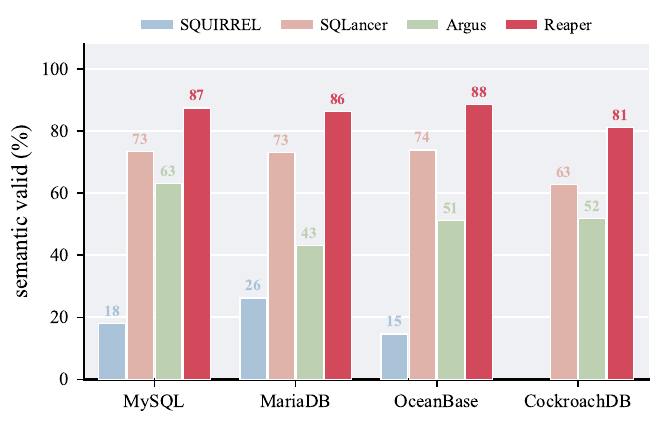}
  \label{fig:sem-valid}
}

\vspace{-2mm}
\caption{Syntactic and semantic validity of generated sibling tests.}
\vspace{-3mm}
\label{fig:sql-validity}
\end{figure}

Manual inspection of the final coverage profiles further shows that
\textsc{Reaper}'s coverage is complementary to, rather than a strict superset
of, the baselines'. While some regions covered by the baselines remain
uncovered by \textsc{Reaper}, it also reaches branches that none of them
exercise. The key difference lies in the dimension along which execution paths are explored: instead of broadly exploring the SQL input space, \textsc{Reaper} replaces the historical construct with its syntactic siblings while preserving the associated bug-triggering context. This changes the exercised handler without discarding the context implicated in the historical bug. As a result, sibling mutation reaches additional implementation paths in components such as optimization, expression resolution, and statement preparation, explaining the complementary coverage observed in the profiles.

\vspace{1mm}
\noindent\textbf{SQL Validity.}
Reaching deeper DBMS logic requires generated queries to remain valid beyond
parsing. Figure~\ref{fig:sql-validity} reports the syntactic validity across
the four evaluated DBMSs and the semantic validity on MySQL, MariaDB, and
OceanBase. \textsc{Reaper} maintains consistently high syntactic validity,
reaching 98\% on MySQL and MariaDB and 97\% on OceanBase and CockroachDB.
\textsc{SQLancer} and \textsc{Argus} are slightly higher at 98--99\%.
The small gap mainly arises from the additional concretization performed by
\textsc{Reaper}: although sibling statement skeletons are derived from the
DBMS grammar, completing sibling-specific operands and context can occasionally
produce queries rejected by the parser. More importantly, \textsc{Reaper} achieves substantially higher semantic
validity, reaching 87\%, 86\%, and 88\% on MySQL, MariaDB, and OceanBase,
respectively, compared with approximately 73\% for \textsc{SQLancer} and
43--63\% for \textsc{Argus}. Mutation-based \textsc{Squirrel} achieves
syntactic validity of 73\%, 84\%, and 66\% and semantic validity of
15--26\% on the three supported DBMSs. These results show that the modest
syntactic cost introduced by concretization is accompanied by a substantial
gain in semantic validity. By incorporating the historical query context,
the Concretizer resolves sibling-specific bindings and operands while
preserving the grammar-defined sibling structure, allowing more generated
tests to exercise the intended sibling handlers.

In summary, \textsc{Reaper} demonstrates stronger bug-finding capability than
the baselines, while also achieving higher branch coverage and semantic
validity with comparable syntactic validity. These results adequately answer
RQ2.

% Keep the two ablation tables together, with separate captions and labels.
\begin{table}[t]
\centering
\footnotesize
\setlength{\tabcolsep}{2.5pt}
\renewcommand{\arraystretch}{1.12}

\begin{minipage}[t]{0.49\linewidth}
\vspace{0pt}
\begin{minipage}[t]{\linewidth}
\caption{Ablation study: unique bugs discovered by each
\textsc{Reaper} configuration.}
\label{tab:ablation}
\end{minipage}
\par\nointerlineskip

\begin{tabularx}{\linewidth}{@{}l*{3}{>{\centering\arraybackslash}X}@{}}
\toprule
\textbf{DBMS}
& \textbf{\textsc{Reaper}}
& \textbf{\textsc{Reaper}\(_{-n_{\mathit{fault}}}\)}
& \textbf{\textsc{Reaper}\(_{-\mathit{conc}}\)} \\
\midrule
MariaDB      & 12 & 0 & 2 \\
MySQL        & 7  & 0 & 0 \\
OceanBase    & 9  & 0 & 1 \\
CockroachDB  & 10 & 1 & 2 \\
\midrule
\textbf{Total} & \textbf{38} & 1 & 5 \\
\bottomrule
\end{tabularx}
\end{minipage}
\hfill
\begin{minipage}[t]{0.49\linewidth}
\vspace{0pt}
\begin{minipage}[t]{\linewidth}
\caption{Ablation of sibling-bug validation on the curated benchmark.}
\label{tab:validation_ablation}
\end{minipage}
\par\nointerlineskip

\begin{tabularx}{\linewidth}{@{}>{\raggedright\arraybackslash}Xrrrr@{}}
\toprule
\textbf{Configuration}
& \textbf{Acc.}
& \textbf{Prec.}
& \textbf{Rec.}
& \textbf{F1} \\
\midrule

LLM-only
& 46.9
& 38.2
& 72.2
& 50.0 \\

Historical Diff + LLM
& 61.2
& 48.6
& 94.4
& 64.2 \\

+ Stage~I (match $P_{\mathit{fail}}$)
& 75.5
& 61.5
& 88.9
& 72.7 \\

+ Stage~II (apply $C_{\mathit{fix}}$)
& 81.6
& 69.6
& 88.9
& 78.0 \\

\midrule

\textbf{Full \textsc{Reaper} (+ $O_{\mathit{fix}}$)}
& \textbf{87.8}
& \textbf{80.0}
& \textbf{88.9}
& \textbf{84.2} \\

\bottomrule
\end{tabularx}
\end{minipage}

\end{table}

\vspace{-2mm}
\subsection{RQ3: Ablation Study}
\label{sec:rq3}

We examine the two key components that enable sibling-bug discovery in
\textsc{Reaper}: the identification of the bug-triggering construct
$n_{fault}$, which guides grammar-guided sibling mutation, and
trigger-constrained concretization, which resolves sibling-specific semantic
requirements while preserving the bug-triggering context recovered from
historical fixes. We compare full \textsc{Reaper} against
\textsc{Reaper}\(_{-n_{\mathit{fault}}}\), which replaces the identified
bug-triggering construct $n_{fault}$ with a randomly selected mutable construct, and
\textsc{Reaper}\(_{-\mathit{conc}}\), which removes the concretizer. 
All configurations use the same historical corpus, target DBMSs, and
validation procedure. We evaluate the contribution of sibling bug validation separately
in \S\ref{sec:rq4}.

\vspace{1mm}
\noindent\textbf{Bug-triggering construct $n_{fault}$.}
The effectiveness of sibling test generation depends first on whether the
bug-triggering construct $n_{fault}$ can be identified reliably. As a direct
LLM baseline, we provide the complete historical fix to a single LLM and ask
it to identify $n_{fault}$ in one pass; this achieves only 63.0\% accuracy.
In contrast, \textsc{Reaper} analyzes the fix from complementary query-side
and repair-side views. The Query Analyzer and Patch Analyzer initially agree
on 78.1\% of the fixes, and the Orchestrator resolves disagreements for another
14.2\%, while the remaining 7.7\% are discarded when the two views remain
inconsistent. Two authors independently inspected a random sample of 50
accepted Bug Profiles and judged the confirmed $n_{fault}$ correct in 94.0\%
of them. These results show that cross-view analysis provides a substantially
more reliable anchor for subsequent sibling testing while filtering
localizations that cannot be consistently established.

The ablation further shows why identifying $n_{fault}$ is important. As shown
in Table~\ref{tab:ablation}, \textsc{Reaper}\(_{-n_{\mathit{fault}}}\) reduces the
number of detected bugs from 38 to 1, with no bug found on MariaDB, MySQL, or
OceanBase. Importantly, this variant still performs the same sibling
mutation, concretization, and validation; the only difference is the
construct selected for mutation. This result demonstrates that
syntactic relatedness alone is insufficient for identifying defect-relevant
sibling handlers. The historical fix provides the semantic context that
connects syntactic siblings with the repaired behavior. Without this
repair context, mutation may reach syntactically similar handlers but lacks
evidence that these handlers are subject to the same repaired requirement.

\vspace{1mm}
\noindent\textbf{Trigger-constrained concretization.}
\textsc{Reaper}\(_{-\mathit{conc}}\) retains the same $n_{fault}$ and therefore
the same defect-relevant sibling candidates, but exposes only 5 bugs compared
with 38 for full \textsc{Reaper}. This contrast separates structural
relevance from executable activation. Grammar-Guided Sibling Mutation can identify a
related handler and construct a corresponding skeleton, but the selected
sibling may introduce new operands, bindings, or execution requirements that
are absent from the historical query. Without concretization, many such
skeletons fail before the behavior of the selected sibling is exercised. Hence,
grammar relations determine \emph{where} to test, but do not by themselves
provide the semantic conditions required to exercise the target handler.

Taken together, the two ablations reveal a division of responsibility in \textsc{Reaper}'s test generation. Identifying the bug-triggering construct $n_{fault}$ concentrates the search on defect-relevant sibling handlers, whereas trigger-constrained concretization enables those handlers to be exercised under a realizable test context. Removing either property substantially degrades bug discovery, showing that sibling testing requires both a bug-relevant target and a semantically valid sibling test.

\subsection{RQ4: Effectiveness of Logic-Bug Validation}
\label{sec:rq4}

To evaluate \textsc{Reaper}'s logic-bug validation capability, we examine how reliably it
distinguishes sibling-bug candidates.  
% from benign projections whose behavior differs from the historical query but remains correct.

\vspace{1mm}
\noindent\textbf{Curated validation benchmark.}
We construct a labeled benchmark to isolate the effectiveness of the
three-stage validation in Section~\ref{sec:oracle}. The benchmark is not
intended to estimate field precision; instead, it provides a fixed set on
which different validation strategies can be compared under identical
candidates. It contains upstream-confirmed sibling bugs, triaged logic
candidates, controlled oracle-replay cases, and manually labeled benign projections.
Each record contains the Bug Profile $\Psi$, the sibling query
$Q'$, its execution outcome, and a human label. Crash and fatal-error
candidates are excluded because this experiment focuses on logic-bug
validation. In total, the benchmark contains 49 logic candidates, including
18 positives and 31 negatives (test cases show different but correct behavior from the fixed historical bug).

\vspace{1mm}
\noindent\textbf{Staged validation effectiveness.}
Table~\ref{tab:validation_ablation} compares configurations that
progressively incorporate historical execution information and the three
components of the Validation Profile: $P_{\mathit{fail}}$, the historical
failure pattern; $C_{\mathit{fix}}$, the fix-derived criterion specifying
expected SQL behavior; and $O_{\mathit{fix}}$, the fix-derived oracle
specifying a checkable condition or relation under the conditions specified in the oracle.
\textbf{LLM-only}, which classifies candidates using their execution
outputs alone, achieves 46.9\% accuracy, 38.2\% precision, 72.2\% recall,
and 50.0\% F1, misclassifying 21 of the 31 benign projections as bugs.
\textbf{Historical Diff + LLM} additionally incorporates the historical
and sibling execution outcomes, increasing recall to 94.4\% and precision
to 48.6\%. It retains 17 of the 18 positive cases but still produces
18 false positives, indicating limited discrimination between sibling
bugs and legitimate execution differences. 
Adding \textbf{Stage~I (match $P_{\mathit{fail}}$)} requires the sibling
execution to exhibit the historical failure pattern. False positives
decrease from 18 to 10, increasing precision to 61.5\% and accuracy to
75.5\%. Recall decreases from 94.4\% to 88.9\%, corresponding to one
fewer retained positive case (17 to 16). The higher recall results from this relatively
permissive decision, which preserves nearly all positive cases while also
admitting many benign projections.
Adding \textbf{Stage~II (apply $C_{\mathit{fix}}$)} checks whether the
fix-derived criterion applies to the selected sibling. This further reduces
false positives to 7 and increases precision to 69.6\% and accuracy to
81.6\%, while recall remains at 88.9\%.
Full \textsc{Reaper} additionally evaluates oracle violation using
$O_{\mathit{fix}}$ under the conditions specified in the oracle, reducing false positives
to 4 and achieving 87.8\% accuracy, 80.0\% precision, 88.9\% recall,
and 84.2\% F1. It retains 16 of the 18 positive cases and correctly
rejects 27 of the 31 benign projections. Following failure-pattern
matching, criterion-applicability and oracle-violation checking together
increase precision from 61.5\% to 80.0\% without reducing recall.
These results support the contribution of the validation profile to
rejecting benign projections while retaining true sibling bugs.

\vspace{-4mm}
\section{Related Work}
\label{5-related}
\noindent\textbf{DBMS Fuzzing.}
DBMS fuzzing has been extensively studied for discovering bugs through
automated SQL generation and mutation. Mutation-based fuzzers, including
Squirrel~\cite{squirrel}, DynSQL~\cite{dynsql}, and
SQLRight~\cite{liang2022detecting}, evolve existing SQL seeds through
AST-level, stateful, or coverage-guided transformations. Generation-based
approaches, such as SQLsmith~\cite{sqlsmith}, Polyglot~\cite{polyglot}, and
SQLancer++~\cite{sqlancerpp}, synthesize SQL queries from grammars or adaptive
generators. Other approaches introduce additional guidance through grammar-free testing, reusable semantic operations, query-plan exploration, coverage-guided traversal, and bug-pattern prioritization [1,4,12,28,29].

Recent DBMS testing approaches have also adopted LLMs to assist SQL generation.
ShQveL~\cite{shqvel} synthesizes SQL fragments to improve dialect coverage,
while Argus~\cite{argus} generates constrained abstract query pairs together
with oracle templates. These approaches improve the diversity and complexity
of generated SQL inputs, but primarily explore the general DBMS input space.
They do not identify a bug-triggering construct from a historical fix and
extend its triggering context to grammar-related sibling handlers.
\textsc{Reaper} instead starts from a historical fix, recovers its Bug
Profile, and uses the bug-triggering construct to direct testing toward
sibling handlers that may contain residual sibling bugs.

\noindent\textbf{Test Oracles for DBMS Testing.}
Reliable DBMS testing requires effective oracles for determining whether an
execution is incorrect. Metamorphic testing approaches, such as
TLP~\cite{tlp} and NoREC~\cite{norec}, construct logically related query
variants to detect inconsistent results. Differential execution approaches,
including DQE~\cite{dqe}, compare multiple executions under related inputs,
while EET~\cite{eet} and CODDTest~\cite{coddtest} exploit expression-level
equivalences. PINOLO~\cite{pinolo} synthesizes approximate metamorphic
relations, and Kangaroo~\cite{kangaroo} combines context-sensitive
instantiation with multi-plan execution. Other approaches derive test oracles
from configuration changes~\cite{mozi}, cross-engine
agreement~\cite{thanos}, or executable SQL
semantics~\cite{liu2024semantic}. These oracle mechanisms rely on predefined relations or specifications,
such as semantic equivalence, differential consistency, configuration
invariance, cross-engine agreement, or reference semantics. However, they
cannot be directly applied to Sibling Bug Validation because execution
differences between sibling and historical tests do not necessarily indicate
a bug.

\textsc{Reaper} instead derives the Validation Profile from the historical
fix. $P_{\mathit{fail}}$ captures the historical failure pattern,
$O_{\mathit{fix}}$ provides the fix-derived oracle, and
$C_{\mathit{fix}}$ provides the fix-derived criterion. During Sibling Bug
Validation, these three components guide the LLM-based validation of whether
the historical failure pattern is present, whether the fix-derived criterion applies to the selected sibling, and whether the observed behavior violates the fix-derived oracle.

\vspace{-2mm}
\section{Conclusion}
\label{6-conclusion}

In this paper, we identified \emph{sibling bugs}, residual DBMS defects that
persist in handlers corresponding to syntactic siblings after a localized fix. We presented
\textsc{Reaper}, a fix-anchored DBMS fuzzing framework that reuses historical
fixes to extend regression testing beyond the originally repaired path.
\textsc{Reaper} recovers Trigger Profile and Validation Profile from historical
fixes, projects bug-triggering queries to syntactic siblings defined by the SQL
grammar, and validates sibling executions using fix-derived evidence.
Evaluated on MariaDB, MySQL, OceanBase, and CockroachDB, \textsc{Reaper}
uncovered 61 previously unknown bugs, of which 57 have been confirmed and 13
have been fixed by developers. These results show that historical fixes can
serve as reusable testing knowledge for systematically exposing residual
defects across handlers corresponding to syntactic siblings.

\vspace{-2mm}
\section{Data Availability}
\label{7-data}
The source code and original evaluation data are hosted anonymously on
Figshare: \url{https://figshare.com/s/a9aab4d572fb108f7ba7}.

% -----------------------------------------------------------------------------
% Data Availability
% -----------------------------------------------------------------------------
% Replace the text below according to the actual status of your artifact.

% -----------------------------------------------------------------------------
% Acknowledgments
% -----------------------------------------------------------------------------
% author identities.
%
% \begin{acks}
% The authors acknowledge ...
% \end{acks}

% -----------------------------------------------------------------------------
% References
% -----------------------------------------------------------------------------

\bibliographystyle{ACM-Reference-Format}
% Avoid an isolated reference continuation before the author-reference table.
\enlargethispage{\baselineskip}
\bibliography{ref}

% Author reference only: keep the unused bug list after all paper content.

\end{document}